\documentclass[twocolumn,twocolappendix,tighten,times]{aastex631}

\usepackage{comment}

\shorttitle{FCC~224}
\shortauthors{Tang et al.}

\begin{document}

\title{An Unexplained Origin for the Unusual Globular Cluster System in the Ultra-diffuse Galaxy FCC~224}

\correspondingauthor{Yimeng Tang}
\email{ymtang@ucsc.edu}

\author[0000-0003-2876-577X]{Yimeng Tang}
\affiliation{Department of Astronomy \& Astrophysics, University of California Santa Cruz, 1156 High Street, Santa Cruz, CA 95064, USA}

\author[0000-0003-2473-0369]{Aaron J.\ Romanowsky}
\affiliation{Department of Astronomy \& Astrophysics, University of California Santa Cruz, 1156 High Street, Santa Cruz, CA 95064, USA}
\affiliation{Department of Physics \& Astronomy, San Jos\'e State University, One Washington Square, San Jose, CA 95192, USA}

\author[0000-0002-2936-7805]{Jonah S.\ Gannon}
\affiliation{Centre for Astrophysics and Supercomputing, Swinburne University, John Street, Hawthorn, VIC 3122, Australia}
\affiliation{ARC Centre of Excellence for All Sky Astrophysics in 3 Dimensions (ASTRO 3D), Australia}

\author[0000-0003-0327-3322]{Steven R.\ Janssens}
\affiliation{Centre for Astrophysics and Supercomputing, Swinburne University, John Street, Hawthorn, VIC 3122, Australia}
\affiliation{ARC Centre of Excellence for All Sky Astrophysics in 3 Dimensions (ASTRO 3D), Australia}

\author[0000-0002-9658-8763]{Jean P.\ Brodie}
\affiliation{Centre for Astrophysics and Supercomputing, Swinburne University, John Street, Hawthorn, VIC 3122, Australia}
\affiliation{ARC Centre of Excellence for All Sky Astrophysics in 3 Dimensions (ASTRO 3D), Australia}
\affiliation{University of California Observatories, 1156 High Street, Santa Cruz, CA 95064, USA} 

\author[0000-0001-9742-3138]{Kevin A.\ Bundy}
\affiliation{Department of Astronomy \& Astrophysics, University of California Santa Cruz, 1156 High Street, Santa Cruz, CA 95064, USA}

\author[0000-0003-3153-8543]{Maria Luisa Buzzo}
\affiliation{Centre for Astrophysics and Supercomputing, Swinburne University, John Street, Hawthorn, VIC 3122, Australia}
\affiliation{ARC Centre of Excellence for All Sky Astrophysics in 3 Dimensions (ASTRO 3D), Australia}
\affiliation{European Southern Observatory, Karl-Schwarzschild-Strasse 2, 85748 Garching bei M\"{u}nchen, Germany}

\author{Enrique A.\ Cabrera}
\affiliation{Department of Physics \& Astronomy, San Jos\'e State University, One Washington Square, San Jose, CA 95192, USA}

\author[0000-0002-1841-2252]{Shany Danieli}
\affiliation{Department of Astrophysical Sciences, Princeton University, Princeton, NJ 08544, USA}

\author[0000-0002-6411-220X]{Anna Ferr\'e-Mateu}
\affiliation{Centre for Astrophysics and Supercomputing, Swinburne University, John Street, Hawthorn, VIC 3122, Australia}
\affiliation{Instituto Astrofisica de Canarias, Av. Via Lactea s/n, E-38205 La Laguna, Spain}
\affiliation{Departamento de Astrofisica, Universidad de La Laguna, E-38200 La Laguna, Tenerife, Spain}

\author[0000-0001-5590-5518]{Duncan A.\ Forbes}
\affiliation{Centre for Astrophysics and Supercomputing, Swinburne University, John Street, Hawthorn, VIC 3122, Australia}
\affiliation{ARC Centre of Excellence for All Sky Astrophysics in 3 Dimensions (ASTRO 3D), Australia}

\author[0000-0002-8282-9888]{Pieter G.\ van Dokkum}
\affiliation{Department of Astronomy, Yale University, New Haven, CT 06511, USA}

\begin{abstract}

We study the quiescent ultra-diffuse galaxy FCC~224 in the Fornax cluster using {\it Hubble Space Telescope} ({\it HST}) imaging, motivated by peculiar properties of its globular cluster (GC) system revealed in shallower imaging. The surface brightness fluctuation distance of FCC~224 measured from {\it HST} is $18.6 \pm 2.7$ Mpc, consistent with the Fornax Cluster distance. We use {\tt Prospector} to infer the stellar population from a combination of multi-wavelength photometry ({\it HST}, ground-based, {\it WISE}) and Keck Cosmic Web Imager spectroscopy. The galaxy has a mass-weighted age of $\sim$ 10 Gyr, metallicity [M/H] of $\sim -1.25$ dex, and a very short formation $e$-folding time of $\tau \sim 0.3$ Gyr. Its 12 candidate GCs exhibit highly homogeneous $g_{\rm 475}-I_{\rm 814}$ colors, merely 0.04 mag bluer than the diffuse starlight, which supports a single burst formation scenario for this galaxy. We confirm a top-heavy GC luminosity function, similar to the two dark matter deficient galaxies NGC~1052-DF2 and DF4. However, FCC~224 differs from those galaxies with relatively small GC sizes of $\sim$~3 pc ($\sim 35\%$ smaller than typical for other dwarfs), and with radial mass segregation in its GC system. We are not yet able to identify a formation scenario to explain all of the GC properties in FCC~224. Follow-up measurements of the dark matter content in FCC~224 will be crucial because of the mix of similarities and differences among FCC~224, DF2, and DF4.

\end{abstract}

\keywords{Dwarf galaxies (416) --- Galaxy formation (595) --- Globular star clusters(656)}

\section{Introduction} \label{sec:intro}

Globular clusters (GCs) are old, tightly bound star clusters associated with all types of galaxies. Observations generally indicate that galaxies share a similar GC magnitude distribution (or GC luminosity function, GCLF), which is close to a Gaussian distribution peaking at $M_V \approx -7.5$ mag (e.g., \citealt{Harris2001,Brodie2006}). Although the turn-over magnitude and width of the GCLF depend on host galaxy mass, this effect is weak \citep{Jordan2006,Villegas2010}. Such a nearly universal peak magnitude has made the GCLF a standard candle for determining the distances of galaxies (e.g., \citealt{Rejkuba2012}).

Owing to the similar mass-to-light ratios of old stellar populations, the observed universal GCLF implies a corresponding log-normal shape of the GC mass function (GCMF). In contrast, newly formed star clusters typically have a power-law initial mass distribution with a truncation at the high-mass end \citep{Larsen2009,Gieles2009,PortegiesZwart2010,Li2017}. After formation, low-mass clusters are preferentially destroyed during the long-term dynamical evolution, due to mass evaporation associated with tidal shocks, dynamical friction, and external tidal fields \citep{Gnedin1997,Fall2001,Vesperini2003,Kruijssen2009,Elmegreen2010}. Dynamical friction can also lead to disruption of the most massive star clusters, especially for low-mass galaxies, which makes the brighter end of the GCLF steeper \citep{Jordan2006}. These effects shape the initial mass distribution of star clusters into the currently observed GCMF. Moreover, the increasing scatter of turnover magnitude for lower-mass galaxies suggests a significant role from variable initial conditions \citep{Jordan2007b,Villegas2010}.

Two notable exceptions to the GCLF universality were identified recently: the GC-rich ultra-diffuse galaxies (UDGs) NGC~1052-DF2 and NGC~1052-DF4 (hereafter DF2 and DF4) associated with the NGC~1052 group, with top-heavy GCLFs that peak $\sim 2$ mag brighter than normal \citep{vanDokkum2018b,vanDokkum2019,Shen2021a}. It was suggested that these anomalies could be explained by overestimation of the distances of the two galaxies \citep{Trujillo2019,Monelli2019}, but this scenario was ruled out by tip of the red giant branch distances measured with very deep {\it HST} imaging \citep{Danieli2020,Shen2021b,Shen2023}. These unusual GCLFs are difficult to explain within the framework of classical models of GC formation and evolution. Coupled with the discovery that DF2 and DF4 both exhibit a deficiency of dark matter (DM; \citealt{vanDokkum2018a,vanDokkum2019,Danieli2019,Shen2023}), it is reasonable to explore the connection between their DM-deficient nature and their overluminous GCs.

So far, two scenarios have been proposed to explain these two unusual properties simultaneously. In the first, high density, high pressure, and highly sheared environments in the early universe produce overmassive GCs, and subsequently intense star formation drives extreme feedback that propels the DM out of the central region \citep{Trujillo-Gomez2021,Trujillo-Gomez2022}. The second scenario is a ``bullet dwarf'' event where two gas-rich galaxies collided at high velocity in a proto-group environment \citep{Silk2019,Shin2020,Lee2021,Lee2024,vanDokkum2022a,Otaki2023}. The collision separates the DM and gas, and overluminous GCs could naturally form in the highly compressed gas. In contrast to these two scenarios, the more conventional scenario for explaining the DM deficiencies through tidal stripping (e.g., \citealt{Ogiya2018,Jackson2021,Moreno2022}) so far does not provide an explanation for the unusual GCs \citep{Ogiya2022a}.

Understanding the evolutionary histories of DF2 and DF4 will be advanced if additional galaxies with similar properties can be found and studied elsewhere. A potentially major step forward is the recent identification of another low surface brightness (LSB) dwarf, FCC~224 in the Fornax cluster, with an apparently top-heavy GCLF like DF2 and DF4 \citep{Romanowsky2024}. Other than the GCLF and diffuse morphology, FCC~224 does not appear remarkable in its photometric properties.

In this work, we make use of new {\it HST} observations and Keck/KCWI spectroscopy of FCC~224 to investigate its GC systems and the galaxy itself in more detail. Our goals are to confirm the overluminous GCs of FCC~224 with higher-quality imaging, and to explore its other properties for evolutionary clues. We organize this paper as follows. In Section \ref{sec:data}, we describe our {\it HST} and Keck/KCWI observations. Section \ref{sec:methods} presents our data analysis methods. We present our results in Section \ref{sec:results} and discuss various evolutionary scenarios in Section \ref{sec:discussion}. A summary follows in Section \ref{sec:conclusion}.

The adopted distance of FCC~224 is the Fornax cluster distance of 20.0 Mpc \citep{Blakeslee2009}, which we will test further in Section~\ref{sec:sbf}. All magnitudes in this work are in the AB system.

\section{Data} \label{sec:data}

\subsection{Photometric data}

\begin{figure}
\centering
\includegraphics[width=0.473\textwidth]{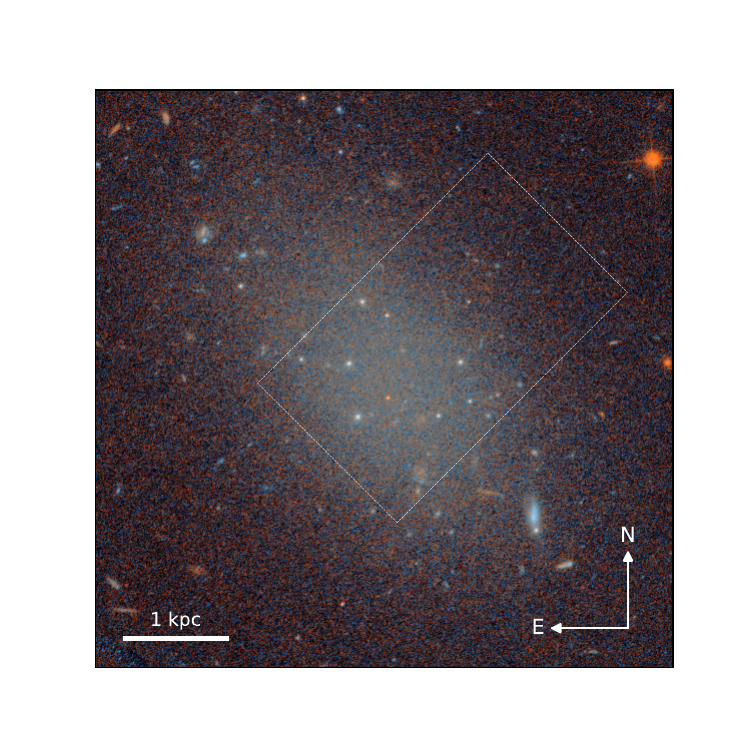}
\caption{{\it HST}/WFC3 F475X/F814W pseudo-color image of the dwarf galaxy FCC~224. The cutout image size is approximately 3 times the major-axis effective radius $R_{\rm e}$ of FCC~224, and the 1 kpc scale bar indicates the physical size. Some bright GCs can be clearly seen superimposed on the galaxy. The white rectangle shows the field of view of the Keck KCWI observation.}
\label{fig:fcc224_hst}
\end{figure}

\begin{figure*}
\centering
\includegraphics[width=1\textwidth]{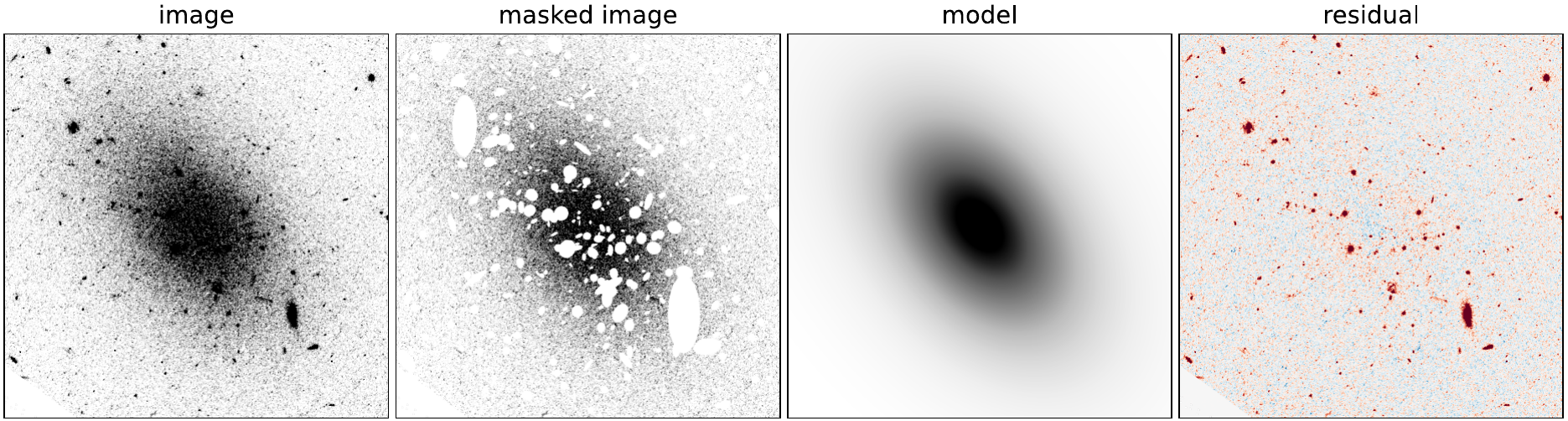}
\caption{FCC~224 $g_{475}$ image as an example of our GALFIT modeling. From left to right, the three panels show the original image, the masked image, the model, and the residuals. The images shown here have sizes of 3 times the major-axis $R_{\rm e}$ of FCC~224, while the actual input images to GALFIT include the entire UVIS chip ($\sim 155 \times 75$ arcsec).}
\label{fig:galfit}
\end{figure*}

Although first cataloged in \cite{Ferguson1989}, \citet{Romanowsky2024} identified FCC~224 as an interesting galaxy using archival imaging from the Dark Energy Camera Legacy Survey (DECaLS; \citealt{Dey2019}) and from {\it HST}/WFPC2. These data were used for a preliminary characterization of the GCLF as ``top-heavy,'' but much higher quality data are needed for confirmation, e.g., the WFPC2 data were heavily affected by cosmic rays.

We observed FCC~224 with the {\it HST} Wide Field Camera 3 (WFC3) using the F475X and F814W filters (hereafter $g_{475}$ and $I_{814}$) with one orbit per filter in the program GO 17149 (PI: A.\ Romanowsky) on 2023 March 3. Four exposures were obtained in each band, but one exposure in $g_{475}$ was discarded because of pointing drifts. The total exposure times were 1791~s in $g_{475}$ and 2372~s in $I_{814}$. We downloaded {\tt flc} individual exposure files from Mikulski Archive for Space Telescopes (MAST) that were calibrated, flat fielded, and corrected for charge transfer efficiency, and combined them with {\tt AstroDrizzle} \citep{Hack2012}. The WFC3 image of FCC~224 is shown in Figure \ref{fig:fcc224_hst}, with pseudo-colors created using the \citet{Lupton2004} algorithm. All the {\it HST} data used in this paper can be found in MAST: \dataset[10.17909/ww65-7p90]{http://dx.doi.org/10.17909/ww65-7p90}.

In addition to {\it HST} imaging, we use additional optical imaging for FCC~224 from DECaLS DR10 in the $g$, $r$, $i$, and $z$ bands. The average seeing for the coadded images in each filter is 1.31, 1.18, 1.28 and 1.12 arcsec, respectively. The 5$\sigma$ depths for point sources in each band are about 26.0, 25.2, 25.3 and 23.9 mag.

For near-infrared (NIR) photometry, we use {\it WISE} imaging \citep{Wright2010} in the W1, W2, W3, and W4 bands (3.4, 4.6, 12.1, and 22.2 $\mu$m). We use the {\tt ICORE} algorithm \citep{Masci2009,Jarrett2012} to create coadded images based on all {\it WISE} exposures within a 15-arcmin diameter region centered on FCC~224. These images are super-sampled with a pixel scale of 1 arcsec, which is a higher resolution than the original pixel size of WISE images (2.75 arcsec per pixel), providing advantages in modeling and subtracting contaminants. With the {\tt ICORE} algorithm, we can keep the full-width at half maximum (FWHM) of the point spread function (PSF) of the coadded images very close to the native {\it WISE} spatial resolution, which is $\sim$~6 arcsec in the W1 and W2 bands.

\begin{figure}
\centering
\includegraphics[width=0.47\textwidth]{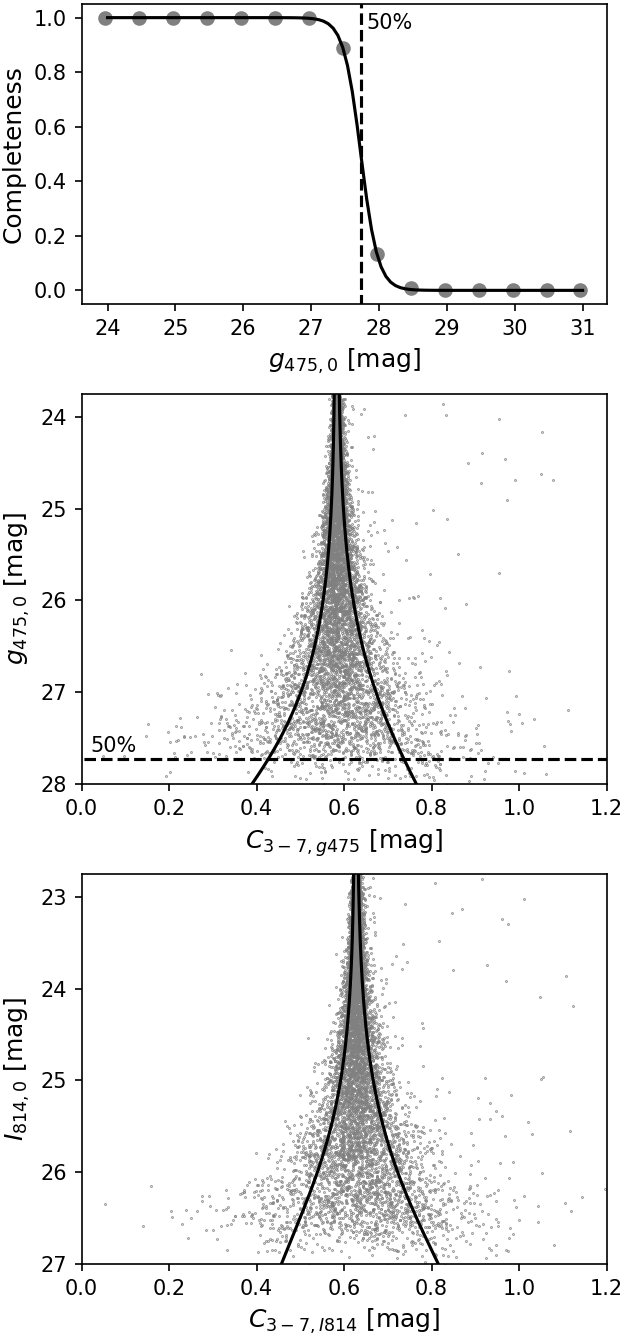}
\caption{{\it Top}: Point-source completeness as a function of apparent magnitude in the $g_{475}$ band, based on detection with {\tt SExtractor} of injected artificial stars. The solid curve is the best-fitted model with Eq.~(\ref{eq:completeness}). The 50\% completeness limit (3-$\sigma$) is $g_{475,0} = 27.7$~mag, marked as a vertical dashed line. {\it Middle and Bottom}: Distribution of the measured concentration parameter $C_{3-7}$ of artificial stars in both bands, defined as the magnitude difference between apertures with 3 and 7 pixel diameters. The solid curves represent the modeling for the 16th and 84th percentiles in the $C_{3-7}$ distribution with Eq.~(\ref{eq:concentration}). The horizontal dashed line is the 50\% completeness limit in $g_{475,0}$.}
\label{fig:completeness}
\end{figure}

\begin{figure*}
\centering
\includegraphics[width=0.95\textwidth]{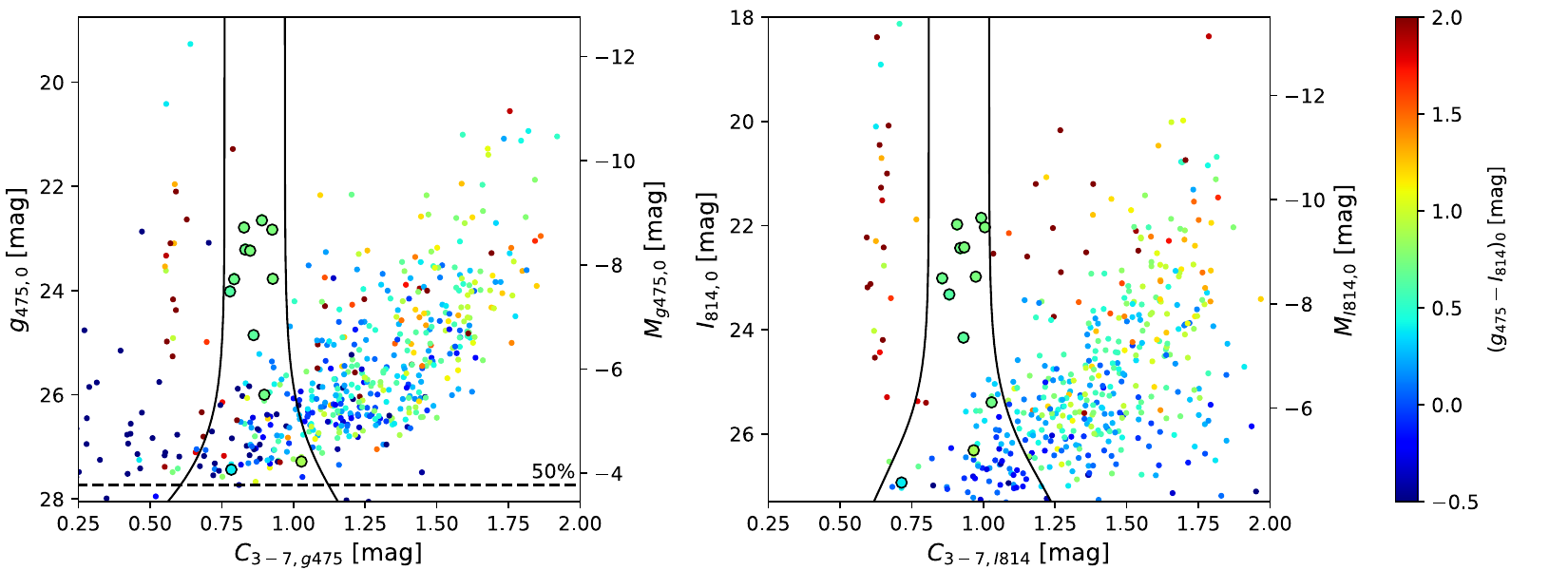}
\caption{Magnitude vs.\ concentration $C_{3-7}$ of the objects detected by {\tt SExtractor}, color-coded by $g_{475}-I_{814}$ color (see colorbar at the right). The solid curves are horizontally shifted from the 16th- and 84th-percentile curves in Figure~\ref{fig:completeness}, with the region in between representing the $C_{3-7}$ range of GC-like objects with sizes from 1 to 5 pc. Objects within the GC $C_{3-7}$ range as well as the GC color range, $0.5<(g_{475}-I_{814})_0 < 1.1$, are selected as GC candidates and circled with black borders. The horizontal dashed line is the 50\% completeness limit in the $g_{475}$ filter. The vertical track of objects at the left shows unresolved sources, which are likely foreground stars. The swath of high-concentration objects toward the right represents background galaxies which are typically larger when brighter. GC candidates fitting both the color and concentration criteria are found mainly at relatively bright magnitudes ($g_{475} \lesssim 25$), while there are few fainter GCs.}
\label{fig:gc_selection}
\end{figure*}

\subsection{Keck/KCWI observation}

Keck Cosmic Web Imager (KCWI; \citealt{Morrissey2018}) spectroscopy of FCC~224 was obtained on 2021 January 13 with one 600s exposure, as part of program U088 (PI: J.~Brodie). The conditions were clear but with poor seeing ($\gtrsim 2.5$ arcsec). The large slicer and BL grating were used, covering a wavelength range of 3553--5576~\AA\ with a spectral resolution of $R \sim$ 900. The data cover an aperture of $\sim$~10~arcsec ($\sim 0.5\ R_{\rm e}$, or $\sim$ 1 kpc at 20 Mpc) in radius for FCC~224, and the sky spectrum was ``on-chip,'' taken from the outer parts of the field of view. 

These data were reduced using the standard KCWI data reduction pipeline, and then were cropped spatially and spectrally to exclude poor data. We extract the spectrum of FCC~224 from a $64 \times 12$ spaxel box ($18.6 \times 16.2$ arcsec) covering the galaxy, and subtract the adjacent area of the datacube as the on-chip sky spectrum. The continuum S/N of the reduced spectrum is $\sim$ 12 \AA${}^{-1}$ at around 4600~\AA. Although we are not able to remove possible contamination sources (including GCs) from the reduced spectrum due to poor seeing, we note that they only contribute a tiny fraction of total flux and should not have a substantial impact on our results.

\section{Methods} \label{sec:methods}

\subsection{Photometric modeling of galaxy light}
\label{sec:galfit}

We use GALFIT \citep{Peng2002} to perform two-dimensional (2D) galaxy modeling to derive the morphological properties and magnitudes of the diffuse starlight of FCC~224. We fit the optical images first, as both {\it HST} and DECaLS are deep enough to constrain the morphological parameters independently. To minimize contamination from the GCs, foreground stars, and background galaxies, we run {\tt SExtractor} \citep{Bertin1996} with low thresholds to generate masks. A combination model is chosen for fitting, consisting of a single S\'ersic model for the galaxy light and a plane sky model for the background, which provides reasonable fits to all of the images. Considering possible morphology variation with wavelength, the fitting of each optical filter is run independently, and all model parameters are allowed to vary, including major-axis effective radius $R_{\rm e}$, S\'ersic index $n$, axis ratio {\it b/a}, and position angle. We build the effective PSF in each band from bright stars using {\tt Photutils}. The fitting region is the entire chip ($\sim 155 \times 75$ arcsec) for the two {\it HST} bands, and is a square region with a side of 8 $R_{\rm e}$ for the DECaLS images. We have tested that our GALFIT fitting is insensitive to the initial parameter guess, the size of the cutout images, and the signal-to-noise threshold for generating the masks.

Comparing the morphological parameters from the optical bands, we find that their differences are smaller than 3\%, suggesting that there are almost no color gradients (radially or azimuthally) in FCC~224. Therefore, we can reasonably assume that in the {\it WISE} bands, the galaxy has the same global morphology. We adopt the morphological parameters from the deepest {\it HST} image, $g_{475}$, in the GALFIT fitting of the {\it WISE} W1 and W2 images, keeping their values fixed, and leaving only the magnitude and background model free. Unlike in the optical imaging, masking of contaminants in {\it WISE} is unsatisfactory because the PSF sizes are very large and have bright and extended wings. Instead, we fit the contaminant sources with PSF models simultaneously with fitting the diffuse starlight, following methods developed by \cite{Tang2024}. Although a small fraction of these sources are background galaxies, they are generally unresolved by {\it WISE} and can be treated as point sources. These contamination sources are detected by {\tt SExtractor} and supplemented by eye, with sources as faint as $\sim 21$ mag in W1 and W2. We also adopt a square fitting region with a size of 8 $R_{\rm e}$ on a side for the {\it WISE} images.

FCC~224 has no detection in the {\it WISE} mid-IR W3 and W4 bands. To estimate the flux upper limits in these two bands, we placed elliptical apertures with random locations and position angles on the W3 and W4 images. The aperture shapes match $R_{\rm e}$ and $b/a$ from the GALFIT fit to the $I_{814}$ image. We set the magnitude limit to correspond to twice the standard deviation of the flux measured in these randomly placed apertures.

GALFIT provides unrealistically small fitting uncertainties, so we recalculate the uncertainties using Monte Carlo simulations, using the same method described in \cite{Tang2024}. As an example of the high quality of the fits, we show the GALFIT best-fitting model and residual for the $I_{814}$ band in Figure \ref{fig:galfit}. We note that there is some mild structure on the residual image, but adding an extra S\'ersic component does not help to increase the fitting quality.

We correct for Galactic extinction with the commonly used dust map by \cite{Schlegel1998}, recalibrated in \cite{Schlafly2011}, and with the extinction curve by \cite{Cardelli1989}. The corrections are 0.032 and 0.016 mag for {\it HST} $g_{475}$ and $I_{814}$ bands, 0.032, 0.023, 0.017 and 0.013 mag for DECaLS $g$, $r$, $i$, $z$ bands, and negligible for {\it WISE}.

\subsection{GC selection}
\label{sec:gc_selection}

Two key characteristics that differentiate GCs from other types of objects are color and size. To determine a reasonable color range of GC candidates, we generate predictions from single stellar population models with ages of 3--14 Gyr and metallicities [M/H] from $-2.0$ to 0 dex, using the Flexible Stellar Population Synthesis code ({\tt FSPS} 3.2; \citealt{Conroy2009}). We adopt the color cut $0.5<(g_{475}-I_{814})_0<1.1$ from these models for GC selection, and require the selected sources to be consistent with this color range within 1-$\sigma$. We note that this color range is consistent with the metal-poor GCs in low-mass early-type galaxies in Fornax and Virgo (e.g., \citealt{Peng2006,Mieske2010}).

To define our photometric selection criteria, we start with artificial star tests. We randomly place 15000 artificial stars (using the PSF generated by {\tt Photutils}), in batches of 100, in the region within 2 $R_{\rm e}$ from the galaxy center (where all the GC candidates are found, see below) on each of the original $g_{475}$ and $I_{814}$ images. The input magnitude ranges are $23.75< g_{475} <31.25$ mag and $22.75< I_{814} <30.25$ mag. We run GALFIT on the newly generated images as described in Section \ref{sec:galfit}, and take the residual images and further clean them by subtracting a smooth background obtained by median filtering. The size of our median filter box is $64\times64$ pix, or $2.5''\times2.5''$, which is big enough to have no effect on the source photometry. Based on these fully cleaned images, we show the recovered fraction with {\tt SExtractor} of artificial stars in different magnitude bins in the top panel of Figure \ref{fig:completeness}, at the 3-$\sigma$ level in $g_{475}$ (the same detection level for the GC identification below). The completeness is modeled by the function
\begin{equation}
\label{eq:completeness}
f(m)=\frac{1}{1+e^{\alpha (m-m_{50})}},
\end{equation}
where $m_{50}$ is the magnitude level at 50\% completeness and $\alpha$ controls the steepness of completeness dropoff toward fainter magnitudes \citep{Harris2016}. The 50\% completeness limit in $g_{475,0}$ is 27.7 mag. In $I_{814,0}$, the 50\% completeness limit is 26.7 mag, which is slightly shallower than $g_{475}$ for our targeted sources with $g_{475}-I_{814} \lesssim 1$.

We use the concentration parameter $C_{3-7}$ as the size indicator, defined by the magnitude difference between apertures of 3 and 7 pixels in diameter (or about 1.6 and 4.0 times the PSF FWHM). In the bottom panels of Figure \ref{fig:completeness}, we plot the $C_{3-7}$ and magnitude distribution in both {\it HST} filters. We model the 16th and 84th percentile curves of the $C_{3-7}$ distribution with the function
\begin{equation}
\label{eq:concentration}
C_{3-7,\ {\rm 16th/84th}}(m)=\frac{p_1}{1+e^{p_2 (m-p_3)}}+p_4,
\end{equation}
where $p_1$, $p_2$, $p_3$ and $p_4$ are free parameters. The model is fitted with artificial stars with $g_{475}<27.5$ or $I_{814}<26.5$ mag, which are bright enough so that the detection level makes no difference in the fitting. These curves represent the 1-$\sigma$ concentration range of point sources. A perfect point source (based on our empirical PSF) has $C_{3-7}$ parameters of 0.58 and 0.63 in the $g_{475}$ and $I_{814}$ bands, respectively. 

At a distance of $\sim$ 20 Mpc, GCs are spatially resolved and more extended than stars, so smaller concentrations and larger $C_{3-7}$ values are expected. We convolve {\it HST} PSFs with circular King models \citep{King1962} with concentration $c=r_t/r_c=100$ and half-light radius steps of $r_h=$~(1, 5) pc, which are reasonable models for GC and are consistent with the known GC candidates in Fornax \citep{Jordan2015}. The $C_{3-7}$ parameters of high-S/N GC models are (0.76, 0.97) and (0.81, 1.02) in the $g_{475}$ and $I_{814}$ filters, respectively. With 1 pc as the adopted lower size limit for our GC selection, the corresponding $C_{3-7}$ boundary comes from shifting the 16th percentile $C_{3-7}$ curves in Figure \ref{fig:completeness} to the right by $0.76-0.58=0.18$ in the $g_{475}$ band, and $0.81-0.63=0.18$ in the $I_{814}$ band. Similarly, 5 pc is set as the upper size limit, and the corresponding $C_{3-7}$ boundaries are from shifting the 84th percentile $C_{3-7}$ curves in Figure \ref{fig:completeness}. These new concentration boundaries are shown in Figure \ref{fig:gc_selection}. We note that the 1-pc lower limit allows for distinguishing point sources from slightly extended sources in a relative sense, as this size is nearly unresolvable for {\it HST} at 20 Mpc (e.g., \citealt{Puzia2014}). We have also tested that if we raise the upper cut to $\sim$8 pc, the GC number remains the same, but with a decreasing ability to distinguish GCs from contaminants.

\begin{figure*}
\centering
\includegraphics[width=0.9\textwidth]{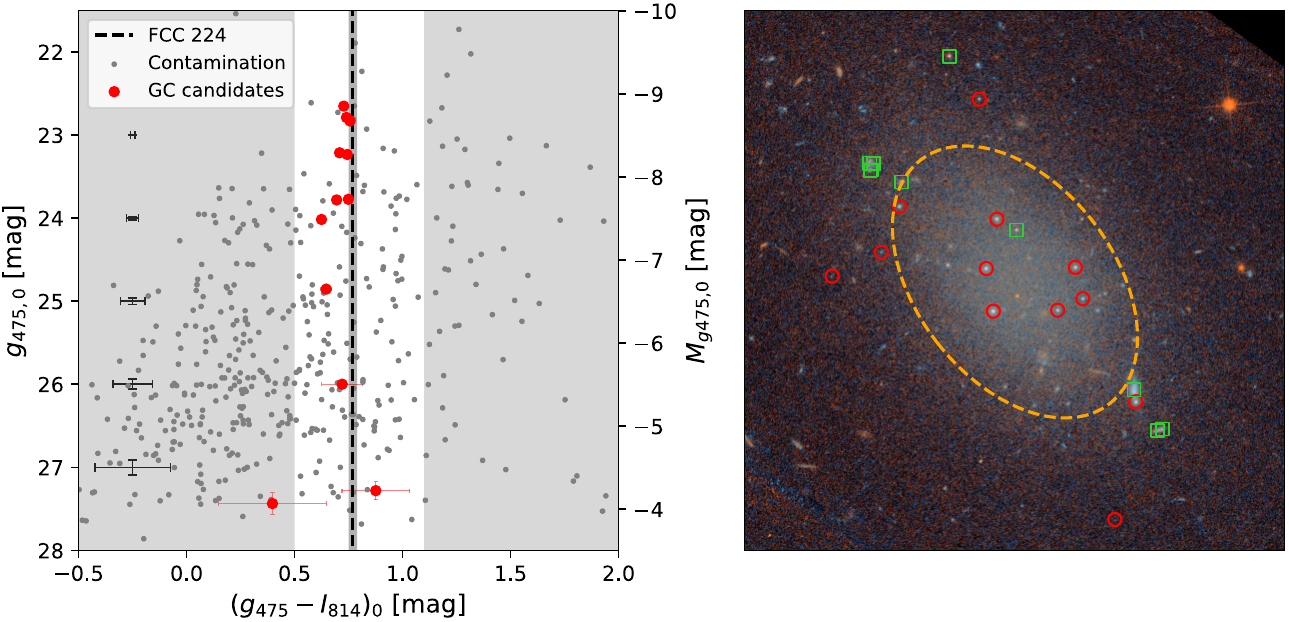}
\caption{{\it Left}: Color--magnitude diagram of objects detected with {\tt SExtractor}, with red circles highlighting GC candidates within 2 $R_{\rm e}$ of the center of FCC~224. Objects in the two gray regions are rejected as GCs based on our color criteria. Gray points that meet the color criteria do not fit the concentration criteria. Example photometric errorbars are shown on the left. The vertical dashed line represents the $g_{475}-I_{814}$ color of the galaxy's diffuse starlight measured with GALFIT (the gray band around the line shows the color uncertainty), which is only slightly ($\sim 0.04$ mag) redder than the 7 brightest GCs. The GC color spread is also only 0.02 mag, suggesting a single-burst formation scenario. {\it Right}: Spatial distribution of GCs, marked with red circles on the pseudo-color image. The orange dashed ellipse represents the isophote at $1 R_{\rm e}$. The five clusters beyond this distance are also the faintest in the sample -- an indicator of mass segregation. The ten bright contaminants listed in Table~\ref{tab:GC_table} are marked with green squares (many of these actually represent different components of background galaxies).}
\label{fig:gc_cmd}
\end{figure*}

\begin{figure}
\centering
\includegraphics[width=0.475\textwidth]{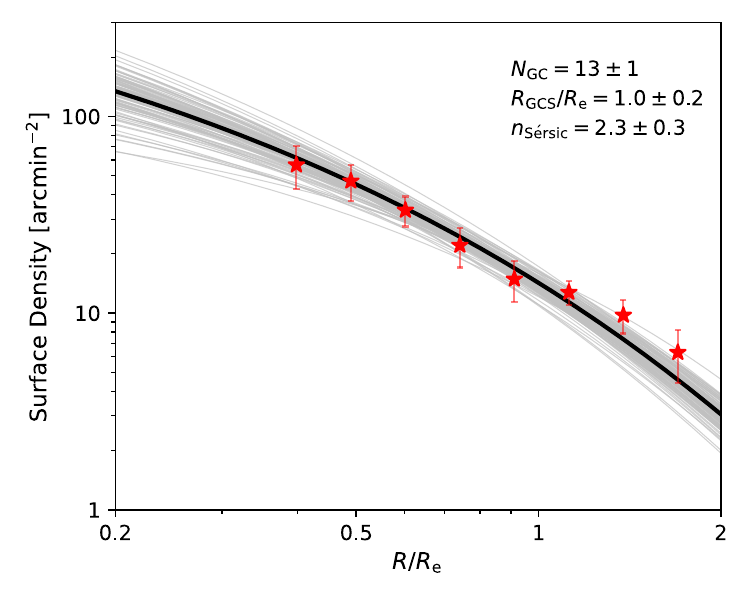}
\caption{The GC surface density profile (red stars), with the uncertainties reflecting the variation of the 100 different samplings of the GC radial distribution with random elliptical annuli. The one-dimensional S\'ersic fit of each random sampling is shown as a light gray curve, and the median of the derived parameters is taken as the best fit (black curve).}
\label{fig:gc_sersic_fitting}
\end{figure}

\begin{deluxetable*}{ccccccc}[htbp]
\label{tab:GC_table}
\tablecaption{Information about the objects identified as FCC~224 GC candidates and as background sources with angular distances of less than 2 $R_{\rm e}$ from the center of FCC~224. We include sky coordinates, apparent and absolute magnitude in $g_{475}$ band, $g_{475}-I_{814}$ color, half-light radius. Colors and magnitudes are extinction-corrected. The list of background sources is truncated at $m_{g_{475},0}=24$.}
\tabletypesize{\small}
\tablecolumns{6}
\tablewidth{20pt}
\tablehead{\colhead{ID} &\colhead{RA (J2000)} & \colhead{Dec (J2000)}
& \colhead{$m_{g_{475},0}$} & \colhead{$M_{g_{475},0}$} & \colhead{$(g_{475}-I_{814})_0$}& \colhead{$r_h$} \\
& [deg] & [deg] & [mag] & [mag] & [mag] & [pc]}
\startdata
GC-1 & 54.88269 & $-$31.80058 & $27.44 \pm 0.13$ & $-4.07 \pm 0.13$ & $0.40 \pm 0.25$ & -- \\
GC-2 & 54.88183 & $-$31.79645 & $22.83 \pm 0.01$ & $-8.67 \pm 0.01$ & $0.76 \pm 0.01$ & $3.2 \pm 0.1$\\
GC-3 & 54.88505 & $-$31.79323 & $23.77 \pm 0.02$ & $-7.73 \pm 0.02$ & $0.75 \pm 0.04$ & $4.8 \pm 0.4$\\
GC-4 & 54.88771 & $-$31.79327 & $22.65 \pm 0.01$ & $-8.85 \pm 0.01$ & $0.73 \pm 0.02$ & $3.8 \pm 0.1$\\
GC-5 & 54.88401 & $-$31.79283 & $24.02 \pm 0.02$ & $-7.49 \pm 0.02$ & $0.62 \pm 0.05$ & $2.4 \pm 0.4$\\
GC-6 & 54.89438 & $-$31.79204 & $27.28 \pm 0.11$ & $-4.22 \pm 0.11$ & $0.88 \pm 0.16$ & -- \\
GC-7 & 54.88800 & $-$31.79177 & $22.79 \pm 0.01$ & $-8.72 \pm 0.01$ & $0.74 \pm 0.02$ & $2.5 \pm 0.5$\\
GC-8 & 54.88432 & $-$31.79172 & $23.21 \pm 0.01$ & $-8.29 \pm 0.01$ & $0.71 \pm 0.03$ & $3.3 \pm 0.2$\\
GC-9 & 54.89234 & $-$31.79120 & $26.00 \pm 0.06$ & $-5.51 \pm 0.06$ & $0.72 \pm 0.09$ & -- \\
GC-10 & 54.88757 & $-$31.79003 & $23.23 \pm 0.01$ & $-8.27 \pm 0.01$ & $0.74 \pm 0.03$ & $3.5 \pm 0.2$\\
GC-11 & 54.89158 & $-$31.78959 & $23.78 \pm 0.02$ & $-7.72 \pm 0.02$ & $0.69 \pm 0.03$ & $2.4 \pm 0.6$\\
GC-12 & 54.88829 & $-$31.78582 & $24.85 \pm 0.03$ & $-6.65 \pm 0.03$ & $0.65 \pm 0.07$ & $3.0 \pm 0.3$\\
\hline
BG-1$^{\rm (e)}$ & 54.88093 & $-$31.79746 & $23.19 \pm 0.01$ & -- & $0.96 \pm 0.03$ & -- \\
BG-2$^{\rm (e)}$ & 54.88073 & $-$31.79740 & $23.22 \pm 0.01$ & -- & $0.35 \pm 0.01$ & -- \\
BG-3$^{\rm (e)}$ & 54.88190 & $-$31.79603 & $21.28 \pm 0.01$ & -- & $0.43 \pm 0.01$ & -- \\
BG-4$^{\rm (p)}$ & 54.88676 & $-$31.79041 & $23.68 \pm 0.02$ & -- & $0.91 \pm 0.03$ & -- \\
BG-5$^{\rm (e)}$ & 54.89151 & $-$31.78873 & $23.64 \pm 0.02$ & -- & $0.15 \pm 0.03$ & -- \\
BG-6$^{\rm (e)}$ & 54.89281 & $-$31.78834 & $23.64 \pm 0.02$ & -- & $0.21 \pm 0.03$ & -- \\
BG-7$^{\rm (e)}$ & 54.89277 & $-$31.78830 & $23.97 \pm 0.02$ & -- & $0.13 \pm 0.03$ & -- \\
BG-8$^{\rm (e)}$ & 54.89267 & $-$31.78809 & $23.90 \pm 0.02$ & -- & $0.40 \pm 0.03$ & -- \\
BG-9$^{\rm (e)}$ & 54.89282 & $-$31.78806 & $22.83 \pm 0.01$ & -- & $1.13 \pm 0.02$ & -- \\
BG-10$^{\rm (p)}$ & 54.88953 & $-$31.78431 & $23.60 \pm 0.02$ & -- & $1.30 \pm 0.03$ & -- \\
... & ... & ... & ... & ... & ... & ... \\
\enddata
\tablecomments{(e) means extended contamination objects, and (p) means point-source like contamination objects.}
\end{deluxetable*}
\vspace{-24pt}

With the two key GC selection criteria determined (i.e., color and concentration), we run {\tt SExtractor} on the fully cleaned $g_{475}$ image to generate source catalogs, with a detection level of 3-$\sigma$. We use {\tt MAG\_AUTO} as the total magnitudes for both bands.

We show the magnitude, concentration $C_{3-7}$, and $g_{475}-I_{814}$ color distributions of the 781 objects identified by {\tt SExtractor} in Figure~\ref{fig:gc_selection}. After applying the $C_{3-7}$ cut of 1 and 5 pc in both bands, and color cut $0.5<(g_{475}-I_{814})_0<1.1$ as mentioned above, we obtain 12 GC candidates. All of them are within twice the major-axis $R_{\rm e}$ from the center of FCC~224. We note that an alternative $C_{3-7}$ upper limit of 10~pc would yield 3 additional, faint GC candidates within $2 R_{\rm e}$, but these would be consistent with the number density of surrounding background objects outside $2 R_{\rm e}$ (2.9 arcmin$^{-2}$).

In Figure \ref{fig:gc_cmd}, we show the color--magnitude diagram (CMD) of the selected GCs and other sources, as well as the GC spatial distribution. We discuss the science interpretation in Sections \ref{sec:gc_position} and \ref{sec:gc_stellar_pop}. To obtain the total number of GCs, we model their radial number density distribution with a S\'ersic profile, following the method described in \cite{Janssens2022} as shown in Figure \ref{fig:gc_sersic_fitting}. We measure the surface density in elliptical annuli with the same position angle as the galaxy, with the radial locations and widths randomly assigned. After creating a suite of 100 surface density profiles with each of them measured from 8 different elliptical annuli, we fit each of the profiles with a one-dimensional S\'ersic profile, using Markov chain Monte Carlo maximum likelihood. Since we do not have any GC candidates beyond $2R_{\rm e}$ and the field of view is not large enough for decent background estimation, we only conduct the fitting within $2R_{\rm e}$ and do not include a background component. The uncertainties of the parameters are calculated from the standard deviations of the 100 fitting results. We find the total GC number of FCC~224 is $13 \pm 1$ (after the magnitude completeness and radial correction, where the latter is dominant), and the half-number radius of the GC radial distribution is $R_{\rm GCS} = 1.0 \pm 0.2\ R_{\rm e}$.

To measure the half-light radius $r_h$ of the selected GCs, we use the PSF-convolved modeling package {\tt ishape} \citep{Larsen1999}. A circular King model with $c=100$ is adopted in fitting to the $g_{475}$ image. We notice that a fixed fitting radius parameter does not provide good fits for all GCs simultaneously. Therefore, we run {\tt ishape} with a range of fitting radii (7, 8, 9, and 10 pixels), and use their likelihood-weighted average as the final measurement value. The results from good fits are not sensitive to the fitting radius, but poor fits can be automatically filtered out due to their large $\chi^2$ values. We next repeat the whole process but switch to a $c=30$ King model, and take the size difference between the results with different concentrations as the uncertainty estimate. Our tests with artificial clusters confirm that this method is unbiased. We obtain reliable size measurements for the brightest 9 GCs with $g_{475,0}<25$ that we have identified. Very similar results are obtained when using the $I_{814}$ image rather than $g_{475}$ for the size measurements.

Photometric information about the selected GCs and bright background sources is listed in Table \ref{tab:GC_table}. We note that independent classification of some of these objects has been obtained through new KCWI spectroscopy (Buzzo et al., in preparation), with redshifts confirming that six of the GCs (ID numbers 3, 4, 5, 7, 8, 10) are indeed GCs, while BG-4 is a star.

\begin{figure*}
\centering
\includegraphics[width=0.9\textwidth]{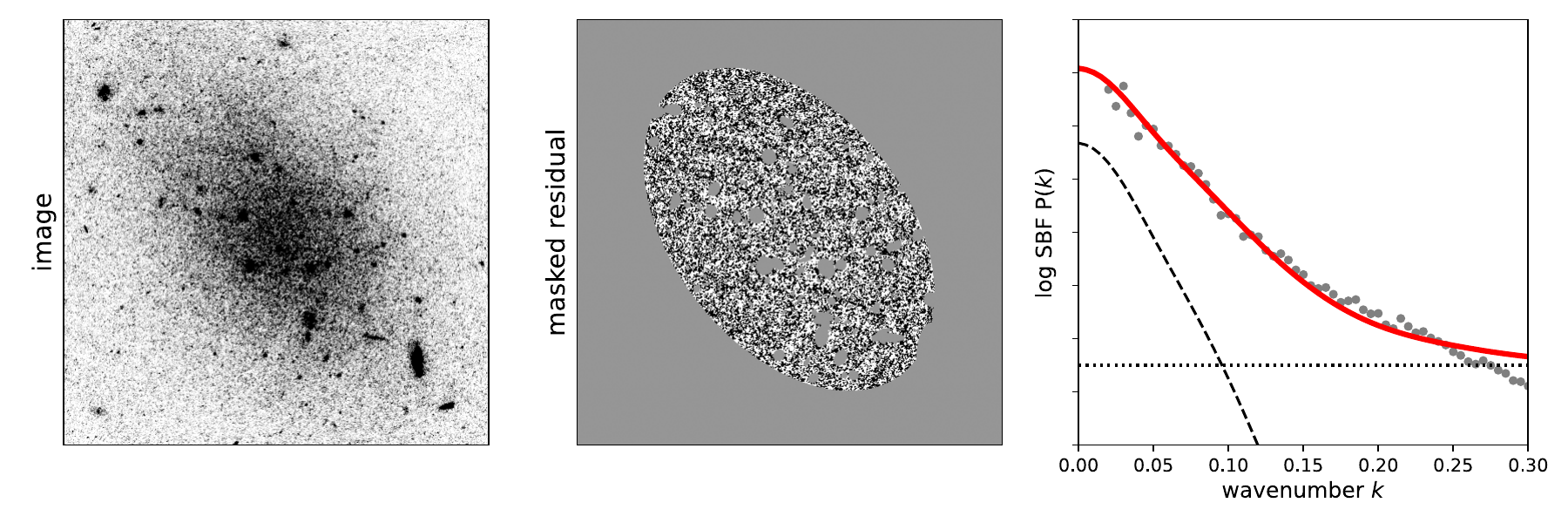}
\caption{The SBF measurement process for FCC~224. The left panel shows the original $I_{814}$ image, and the middle panel is the normalized residual image. The contamination sources and the region outside $1 R_{\rm e}$ are masked. The right panel shows the azimuthally averaged power spectrum (gray points) and the best-fitting combination of the PSF and white-noise components (red solid curve). The two components are also shown individually with the black dashed and dotted lines, respectively. The SBF distance $18.6 \pm 2.7$ Mpc is consistent with the Fornax cluster distance of 20.0 Mpc \citep{Blakeslee2009}.}
\label{fig:sbf}
\end{figure*}

\begin{figure*}
\centering
\includegraphics[width=1\textwidth]{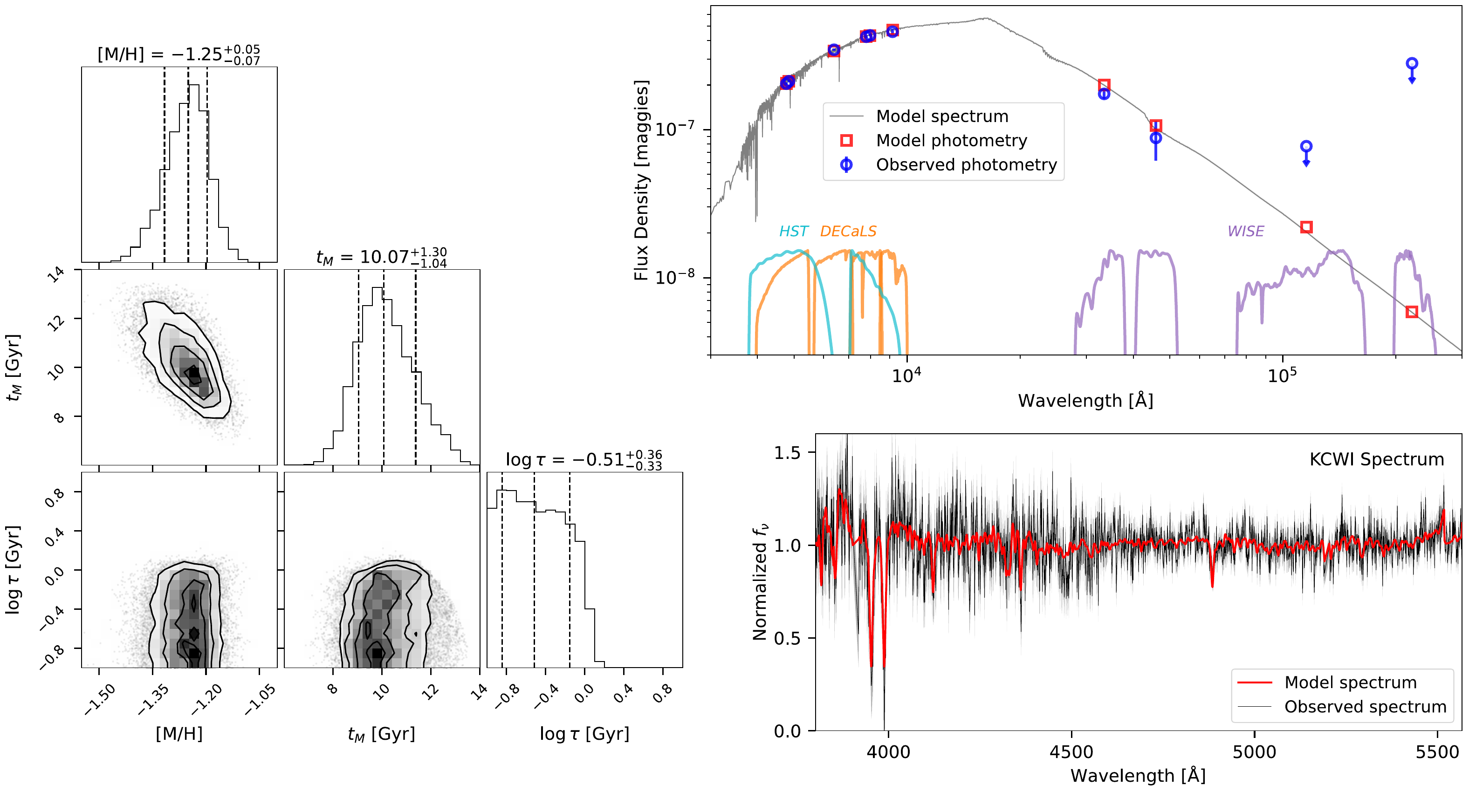}
\caption{The stellar population inference for FCC~224 from combined photometric and spectroscopic fitting with {\tt PROSPECTOR}. The upper right panel shows the observed photometry (blue circles), model photometry (red squares), and model spectrum (gray curve). The upper limits of {\it WISE} W3 and W4 are 1-$\sigma$ limits. At the bottom of this panel, we plot the bandpasses of all the filters used in the fitting, including {\it HST}, DECaLS, and {\it WISE}. The lower right panel shows the observed KCWI spectrum (black) and model spectrum (red). The left panels show one-dimensional (histograms) and two-dimensional (contours and shaded regions) projections of the posterior probability distribution function for three key parameters of the stellar populations (metallicity, age, star formation decay time). On the top of each histogram is the median value of the posterior for that corresponding parameter, with the errorbars giving the 16th and 84th percentiles. We find that FCC~224 has an old and metal-poor stellar population, with a single burst of star formation (a very small $\tau$ of $\sim 0.3$ Gyr).}
\label{fig:prospector}
\end{figure*}

\subsection{Distance measurement}
\label{sec:sbf}

We determine the line-of-sight distance of FCC~224 based on the surface brightness fluctuation (SBF) signal from the $I_{814}$ image. As mentioned in Section \ref{sec:gc_selection}, a smooth model is subtracted from the original image and the net fluxes from the galaxy and sky are fully cleaned. We normalize the residual image by dividing by the square root of the smooth model. The contamination sources are masked with {\tt SExtractor}, and we only use the region within the 1 $R_{\rm e}$ ellipse to achieve higher S/N for our SBF measurement, with the masked residual image shown in the middle panel of Figure \ref{fig:sbf}. Then, we measure the SBF by Fourier transforming the residual image and calculating the azimuthally averaged power spectrum, presented in the right panel of Figure~\ref{fig:sbf}. The one-dimensional power spectrum of the PSF is obtained in the same way. 

\cite{Tonry1988} gave a relation between the residual image power spectrum and the PSF power spectrum:
$$P_{\rm image}(k)=P_0 \times P_{\rm PSF}(k)+P_1,$$
where $P_0$ is the SBF signal to be measured, and $P_1$ is the white-noise term. We fit the power spectrum to get the two parameters $P_0$ and $P_1$, and find the SBF magnitude $\bar{m}_{I_{814}}={\rm ZP}_{I_{814}}-2.5\log(P_0/N)=29.74 \pm 0.03$, where $N$ is the number of unmasked pixels in the middle panel of Figure \ref{fig:sbf}. We fit only over the wavenumber range of $0.03<k<0.30$, as the smaller and larger wavenumbers are dominated by residual large-scale structure in the image and by noise correlations, respectively \citep{Mei2005}. \cite{Carlsten2019} calibrated the absolute SBF magnitude empirically for low mass galaxies, but in the CFHT (Canada–France–Hawaii Telescope) filter system. Using {\tt FSPS}, the {\it HST} $I_{814}$ SBF magnitude and $g_{475}-I_{814}$ color of FCC~224 can be transformed to the CFHT system: $\bar{m}_i=29.77$ and $g-i=0.74$. According to Eq. (4) in \cite{Carlsten2019}, the absolute SBF magnitude in the CFHT $i$-band is $\bar{M_i}=-1.58 \pm 0.32$, and therefore the SBF distance of FCC~224 is $18.6 \pm 2.7$ Mpc. This large distance uncertainty arises mainly from the calibration systematic uncertainty. We note that if using the \cite{Greco2021} theoretical calibration or the \cite{Kim2021} calibration from observations of low-luminosity galaxies, we find very similar results of 18.5 Mpc and $18.8 \pm 1.4$ Mpc, respectively. Since our SBF distance agrees with the Fornax cluster distance of 20.0 Mpc \citep{Blakeslee2009} within the uncertainties, and since the radial velocity of FCC~224 is almost the same as the average velocity of the Fornax cluster (see below), we adopt a 20.0 Mpc distance by default for the rest of this paper. There are no qualitative changes to the conclusions of this paper if 18.6 Mpc is used instead.

\subsection{Stellar population modeling}

We analyze the stellar population of the diffuse starlight of FCC~224 using the Bayesian inference code \texttt{Prospector} (1.2.0; \citealt{Johnson2021}), which allows us to model the photometric and spectroscopic data simultaneously. Our stellar components are modeled with the MIST isochrones \citep{Choi2016} by {\tt FSPS}, based on the MILES stellar spectral library \citep{Falcon-Barroso2011}\footnote{We note that the chemical makeup of the MIST isochrones is not fully consistent with the MILES spectral library, which may introduce non-linear biases to the age and metallicity estimates.}. We adopt the \cite{Kroupa2001} initial mass function. Here we enforce a no-dust condition, as red quiescent galaxies typically have little to no dust. Even when we allow dust to vary in the fitting, the posterior distribution still indicates almost no dust. A simple delayed-tau parametric star formation history (SFH) model is assumed, wherein the star formation rate rapidly rises at the beginning and then decreases exponentially over time. \cite{Greco2018} found low surface brightness galaxies can be well modeled with an exponentially declining SFH. Although the fitting results might depend on the adopted form of the SFH, we find that some key parameters, such as stellar mass, mass-weighted age, and metallicity, are not sensitive to it.

We set linearly uniform priors on all of our seven free parameters in the SED fitting, including redshift ($0 < z < 0.01$), log stellar mass ($7 < \log (M_\star/M_{\odot}) < 9$), metallicity ($-2 < {\rm [M/H]} < 0$), star formation start time ($1.0 < t_{\rm age} < 13.8$ Gyr), and log $e$-folding timescale ($0.1 < \tau < 10$ Gyr). To match the spectral resolution between our KCWI spectrum and the MILES templates, we set the {\tt sigma\_smooth} parameter fixed at 125 km~s$^{-1}$. A free twentieth-order Chebyshev polynomial is included to calibrate the difference in continuum shapes between photometric and spectroscopic data. We use the dynamic nested sampling algorithm \texttt{dynesty} \citep{Speagle2020} to sample the posteriors.

Using the two parameters $t_{\rm age}$ and $\tau$ in the delayed-tau SFH model, we calculate the mass-weighted age $t_M$. The results of our stellar population modeling are shown in Figure \ref{fig:prospector}. The radial velocity of FCC~224 is $1422 \pm 13$ km~s$^{-1}$, given by {\tt Prospector}. As a comparison, the mean recessional velocity of the Fornax cluster is 1454 km~s$^{-1}$, with a velocity dispersion of 286 km~s$^{-1}$ \citep{Maddox2019}. Both our distance and velocity measurements are consistent with FCC~224’s association with the Fornax cluster.

\section{Results} \label{sec:results}

\subsection{FCC~224 galaxy properties}

\begin{deluxetable}{lc}
\tabletypesize{\small}
\tablecolumns{2}
\tablehead{
\colhead{Properties} 
&\colhead{Value}
}

\startdata
RA (J2000) [deg] & 54.88683\\
Dec (J2000) [deg] & $-31.79218$\\
adopted distance [Mpc] & 20.0 \\
$R_{\rm e}$ [arcsec] & $19.5 \pm 0.2$ \\
$R_{\rm e}$ [kpc] & $1.89 \pm 0.01$ \\
$n$ & $0.75 \pm 0.02$ \\
$b/a$ & $0.64 \pm 0.01$ \\
P.A. [deg] & $37.8 \pm 0.5$ \\
$m_{I_{814},0}$ [mag] & $15.91 \pm 0.02$ \\
$M_{I_{814},0}$ [mag] & $-15.60 \pm 0.02$ \\
$(g_{475}-I_{814})_0$ [mag] & $0.77 \pm 0.02$ \\
$(g-i)_{\rm DECaLS,0}$ [mag] & $0.79 \pm 0.03$ \\
$\mu_{g,0}$ [mag arcsec${}^{-2}$] & $23.97 \pm 0.03$ \\
$\langle \mu_{g} \rangle_{\rm e}$ [mag arcsec${}^{-2}$] & $24.69 \pm 0.03$ \\
radial velocity [km~s$^{-1}$] & $1422 \pm 13$ \\
SBF distance [Mpc] & $18.6 \pm 2.7$ \\
log $M_*/M_{\odot}$  & $8.24^{+0.04}_{-0.03}$ \\
stellar age [Gyr] & $10.1^{+1.3}_{-1.0}$ \\ 
stellar metallicity [M/H] [dex] & $-1.25^{+0.05}_{-0.07}$ \\
log $\tau$ [Gyr] & $-0.51^{+0.36}_{-0.33}$ \\
$N_{\rm GC}$ & $13 \pm 1$ \\
$R_{\rm GCS}/R_{\rm e}$ & $1.0 \pm 0.2$ \\
$M_{\rm GCS}/M_{\odot}$ & $(3.6 \pm 0.3) \times 10^6$\\
\enddata

\caption{Various properties of FCC~224, including adopted distance, major axis effective radius $R_{\rm e}$ in arcsec and kpc, S\'ersic index $n$, axis ratio $b/a$, position angle, total magnitude in the $I_{814}$ band, $(g_{475}-I_{814})_0$ color, $(g-i)_{\rm DECaLS,0}$ color, central surface brightness in the $g$-band $\mu_{g,0}$, average surface brightness $\langle \mu_{g} \rangle_{\rm e}$, radial velocity, SBF distance, total stellar mass $M_*$, mass-weighted stellar age and metallicity, star formation $e$-folding time $\tau$, GC number $N_{\rm GC}$, GC system half-number radius $R_{\rm GCS}$, and total GC system mass $M_{\rm GCS}$. All magnitudes and colors are extinction-corrected, and the distance uncertainty is not included in the uncertainties in effective radius, absolute magnitude, and stellar mass.}
\label{tab:properties}
\end{deluxetable}
\vspace{-10pt}

\begin{figure}
\centering
\includegraphics[width=0.472\textwidth]{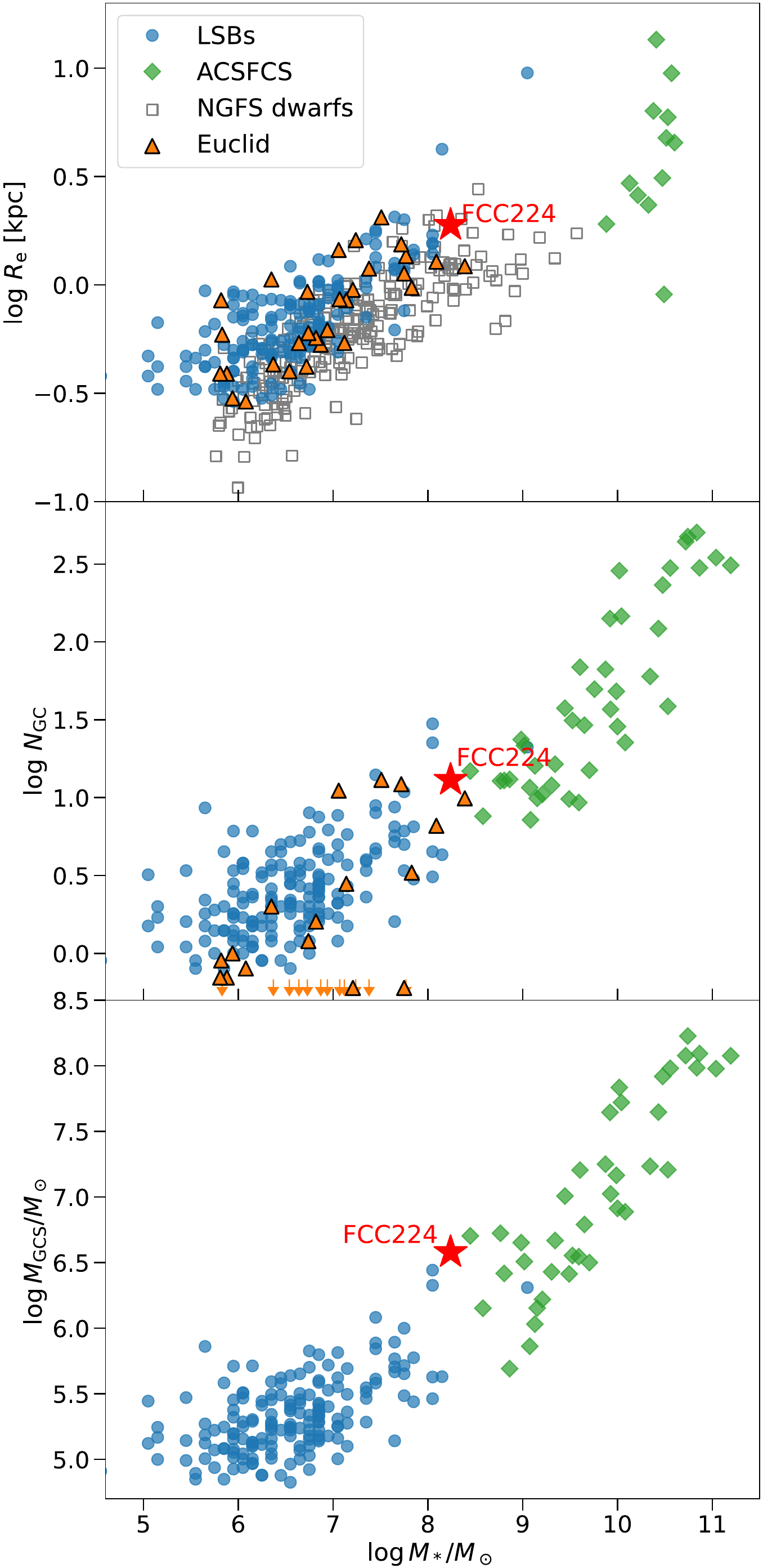}
\vspace{-20pt}
\caption{Trends of Fornax galaxy properties with stellar mass, showing LSB galaxies (blue circles), NGFS dwarfs (gray open squares), ACSFCS galaxies (green diamonds), and {\it Euclid} dwarfs (orange triangles). FCC~224 is shown in all three panels as a red star. {\it Top}: Semi-major axis effective radius. The lower-mass ACSFCS galaxies are missing here because their $R_{\rm e}$ values were not provided in \citet{Liu2019}. {\it Middle and bottom}: GC number and system mass, respectively (not available for NGFS dwarfs). The downward pointing arrows represent the {\it Euclid} dwarfs with $N_{\rm GC}=0$. Compared to other dwarfs, FCC~224 seems relatively large in size, normal in GC number, and has a relatively high total GC system mass.}
\label{fig:gc_compare_with_fornax_lsb}
\end{figure}

We present various properties of FCC~224 in Table \ref{tab:properties}. FCC~224 has a low central surface brightness ($\mu_{g,0}= 24.0$ mag arcsec${}^{-2}$) and a large size ($R_{\rm e}=1.9$ kpc)\footnote{\cite{Tanoglidis2021} reported a much smaller $R_{\rm e}$ and higher surface brightness. We note that their modeling appears to have been challenging for FCC~224 in particular, with large reduced $\chi^2>3$ in all filters.}, fitting the definition of UDGs from \citet{vanDokkum2015}. The isophotes are close to standard ellipses, with almost no diskiness, boxiness, or other asymmetries observed. The top panel of Figure \ref{fig:gc_compare_with_fornax_lsb} shows the size--mass distribution of other Fornax galaxies for comparison. These include LSB dwarfs from Fornax Deep Survey VST/OmegaCAM imaging \citep{Prole2019}, dwarfs from the Next Generation Fornax Survey (NGFS; \citealt{Eigenthaler2018}), galaxies from the ACS Fornax Cluster Survey (ACSFCS; \citealt{Jordan2007a,Liu2019}), and dwarfs observed with {\it Euclid} \citep{Saifollahi2024}. Note that there is some overlap between these samples, and thus some points are duplicates. We see that FCC~224 is more diffuse than the majority of other Fornax dwarfs, but is not extreme. Its size and stellar mass are very similar to those of DF2 and DF4 \citep{Tang2024}. Otherwise, the global properties of FCC~224 align well with the typical range of Fornax dwarfs \citep{Eigenthaler2018,Venhola2022,Chamba2024}.

Our {\tt Prospector} fitting of FCC~224 photometric and spectroscopic data yields a mass-weighted stellar age of 10 Gyr, a metallicity [M/H] of $-1.25$ dex, and a total stellar mass of $1.7\times 10^8\ M_{\odot}$. The best-fit SFH resembles a single burst formation scenario, with an $e$-folding timescale of only 0.3 Gyr. These values are fairly consistent with results for other GC-rich UDGs in the literature (e.g., \citealt{FerreMateu2023,Buzzo2024}).

We measure the $g_{475}-I_{814}$ color radial profile of FCC~224 and find a color gradient of $-0.01 \pm 0.06$ mag dex${}^{-1}$ out to $\sim 2R_{\rm e}$. This is consistent with no gradient and suggests that FCC~224 could have a uniform stellar population at all radii, with almost the same age and metallicity -- providing extra support for the single burst formation scenario. As a comparison, \cite{Koleva2011} found that Fornax dwarfs with $M_B \gtrsim -17$ have average age and metallicity gradients of 0.18 dex dex$^{-1}$ and $-0.2$ dex dex$^{-1}$, respectively. \cite{Bidaran2023} obtained an age gradient of $\sim 0$ and [M/H] gradient of $\sim -0.2$ dex dex$^{-1}$ on average for nine Virgo dwarf ellipticals with $M_V \sim -17.5$ mag. All the evidence here suggests that FCC~224 may not be on a typical evolutionary pathway of dwarf galaxies. Similar to FCC~224, DF2 and DF4 do not show significant color gradients \citep{Tang2024}.

\begin{figure}
\centering
\includegraphics[width=0.473\textwidth]{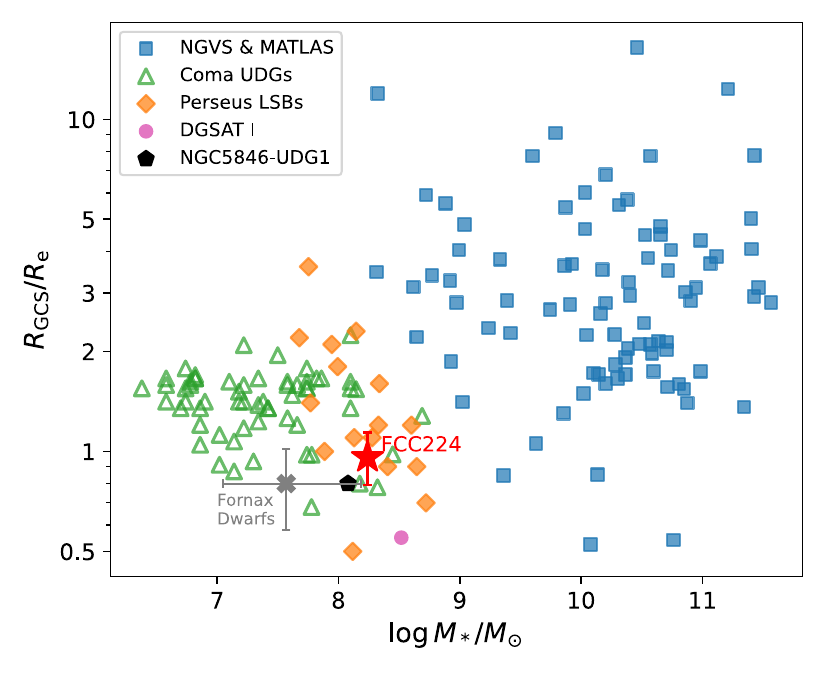}
\caption{The ratio between half-light radii of GC system and host galaxy, vs.\ stellar mass. Blue squares show data from the Next Generation Virgo Cluster Survey (NVGS) and MATLAS Survey, green triangles show Coma cluster UDGs, orange diamonds show Perseus cluster LSB dwarfs, the pink circle shows the ``field'' UDG DGSAT~I, and the black pentagon shows the group UDG NGC~5846-UDG1. The gray cross for the Fornax dwarfs is the measurements of the stacked GC number radial profiles. The horizontal errorbars represent the 16th to the 84th percentiles of the stellar mass distribution of these dwarfs. FCC~224 is overplotted as the red star.}
\label{fig:rgc_re_mass}
\end{figure}

\subsection{GC system number and mass}
\label{sec:gc_number_mass}

We present various properties of the GC system of FCC~224 in this and the following subsections. As mentioned in Section \ref{sec:gc_selection}, the total GC number of FCC~224 is $13 \pm 1$. Compared to other Fornax dwarfs in the literature (middle panel of Figure \ref{fig:gc_compare_with_fornax_lsb}; \citealt{Prole2019,Liu2019,Saifollahi2024}), FCC~224 seems to follow the standard $N_{\rm GC}$--$M_*$ relation. Our {\tt Prospector} fitting gives a prediction for the FCC~224 $V$-band absolute magnitude of $M_V = -15.06$, and therefore the implied GC specific frequency is $12.3 \pm 1.2$, defined as $S_N=N_{\rm GC} \times 10^{0.4(M_V+15)}$ \citep{Harris1981}. Such a specific frequency is close to the typical value for the Fornax dwarfs of similar mass studied by \cite{Liu2019} and \cite{Prole2019}.

Given the unusual GCLF of FCC~224 (discussed later), we will not base the total mass of the GC system, $M_{\rm GCS}$, on the GC numbers, but instead add up the individual GC masses. We find a total GC luminosity of $(2.00 \pm 0.01) \times 10^6\ L_{\odot}$ in the $g_{475}$ band, and need to multiply this by a stellar mass-to-light ratio $M/L$. Given the small color difference between the GCs and diffuse starlight (see Section \ref{sec:gc_stellar_pop}), we can reasonably use the galaxy $M/L$ in estimating the GC stellar masses. From our {\tt Prospector} fitting of the galaxy light, we find $M/L = 1.87_{-0.11}^{+0.16} M_\odot/L_\odot$ in the $g_{475}$ band, so the total GC mass is $M_{\rm GCS} = (3.8 \pm 0.3) \times 10^6\ M_{\odot}$, which is $\sim$ 2\% of the stellar mass of FCC~224. The brightest GC has a mass of about $7 \times 10^5\ M_{\odot}$, and the mean mass is $3 \times 10^5 M_\odot$. As a comparison, the Milky Way has a mean GC mass of $3 \times 10^5 M_\odot$ \citep{Baumgardt2018}, which is similar to FCC~224, while the expectation for a normal dwarf with the stellar mass of FCC~224 is only $10^5 M_\odot$ \citep{Harris2017}.

\begin{figure}
\centering
\includegraphics[width=0.473\textwidth]{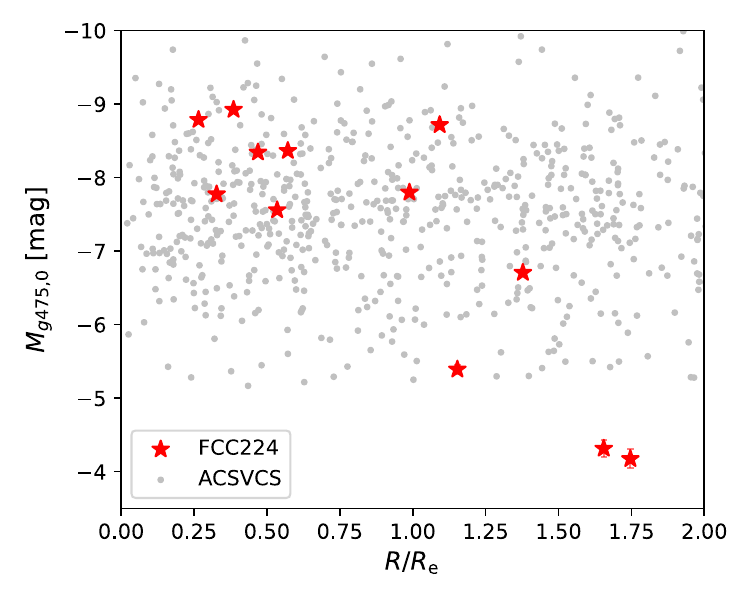}
\caption{The absolute magnitude of GCs in the $g_{475}$ band vs.\ their radial distance from the galaxy center. Mass segregation in FCC~224 (red stars) is evident, with the brighter and fainter GCs located at smaller and larger radii, respectively. A comparison sample from low-mass Virgo cluster galaxies (ACSVCS; gray dots) shows no such trend.}
\label{fig:mass_segregation}
\end{figure}

In the bottom panel of Figure \ref{fig:gc_compare_with_fornax_lsb}, we compare FCC~224 to the $M_{\rm GCS}$--$M_*$ distribution of the other Fornax galaxies. For the Fornax LSBs in \cite{Prole2019}, we derive the GC system masses from the GC numbers, while adopting an average GC mass dependence on host galaxy luminosity from \citet{Harris2017}, and assuming that the GCs and host galaxies have the same $M/L$ in the $V$-band. The GC system masses of ACSFCS galaxies are estimated by using the GC numbers in \cite{Liu2019} and the GCLF parameters in \cite{Villegas2010}, while assuming the GCs have the same $M/L$ in the ACS $z_{850}$ band as their host galaxies. These $M/L$ assumptions likely overestimate the masses of the GCs since their colors are often bluer than their host galaxies (but not in the case of FCC~224). After making all these calculations, we see in Figure~\ref{fig:gc_compare_with_fornax_lsb} that FCC~224 has a more massive GC system than expected at its mass. This suggests that the FCC~224 GCs are overmassive on average, since we have not found that FCC~224 has significantly more GCs than Fornax galaxies of similar stellar masses.

\subsection{GC spatial distribution}
\label{sec:gc_position}

The half-number radius of the radial distribution of the FCC 224 GC system is $R_{\rm GCS} = 1.0 \pm 0.2\ R_{\rm e}$, as calculated in Section \ref{sec:gc_selection}. Unlike the previous properties, there are not individual $R_{\rm GCS}$ values available for Fornax dwarfs for direct comparisons. Instead, we compare to galaxies from the literature in other environments \citep{Amorisco2018,Saifollahi2022,Janssens2022,Janssens2024}, mostly galaxy clusters, plotting $R_{\rm GCS}/R_{\rm e}$ versus host galaxy stellar mass in Figure \ref{fig:rgc_re_mass}. Notably, \citet{Lim2024} carried out a landmark synthesis of GC system sizes spanning both cluster (Virgo) and group (MATLAS; \citealt{Habas2020}) environments, from giant galaxies down to high-mass dwarfs. They reported an average value of $R_{\rm GCS}/R_{\rm e} \sim$~2--4  for these galaxies, with considerable scatter, and higher values on average in lower density environments.  There was also a hint that GC systems become relatively compact at lower masses ($\log_{10}(M_\star/M_\odot) \sim 8.5)$, with $R_{\rm GCS}/R_{\rm e} \sim 1$. With our inclusion of lower-mass dwarfs in Figure~\ref{fig:rgc_re_mass} (the first time such a large dataset has been plotted), this trend appears to be confirmed, with $R_{\rm GCS}/R_{\rm e} \sim 1.5$ on average for the mass range of $\log_{10}(M_\star/M_\odot) \sim$~6.5--8.5. The caveat here is that all of the lower-mass data are for dwarfs selected to be relatively large (mostly UDGs), and therefore this could be a biased trend. 

The one source of information about Fornax dwarf GC system sizes is from {\it Euclid}, using a stacked analysis of 14 galaxies from a mass range of $\log_{10}(M_\star/M_\odot) \sim$~7--8.3 \citep{Saifollahi2024}, which was from a core region of the cluster not including FCC~224. As shown in Figure \ref{fig:rgc_re_mass}, they found an overall $R_{\rm GCS}/R_{\rm e} \sim 0.8$, which seems to suggest more compact GC systems in Fornax than in Coma and Perseus, with the caveat that there are different selection effects and methodologies in these results. 

In this context, the GC system of FCC~224 appears to have a normal radial distribution. There are no published $R_{\rm GCS}$ measurements for DF2 and DF4, but using the catalogs of bright GCs from \citet{Shen2021a} and the galaxy sizes from \citet{Tang2024}, we use median distances of the redshift-confirmed GCs in each system to estimate $R_{\rm GCS}/R_{\rm e} \sim$~1.2 and 2.9, respectively. These estimates do not include faint GCs, whose inclusion could change the $R_{\rm GCS}/R_{\rm e}$ values, especially since there may be mass segregation. These galaxies have very similar stellar masses to FCC~224 \citep{Tang2024}, and their GC systems have average and large extents, respectively. Another UDG discovered having a top-heavy GCLF, DGSAT~I \citep{Janssens2022}, shows a much more compact configuration with $R_{\rm GCS}/R_{\rm e}=0.67 \pm 0.03$.

One additional complexity to the GC spatial distribution is evident in the right panel of Figure~\ref{fig:gc_cmd}, where the faint and bright GCs of FCC~224 are all found at larger and smaller radii, respectively. This {\it mass segregation} is further illustrated by Figure \ref{fig:mass_segregation}, where we see that the eight brighter and four fainter GCs (with $M_V \sim -7$ as the boundary) are all found inside and outside of $1.1 R_{\rm e}$, respectively. Among the bright GCs, there is no further sign of mass segregation, and for these GCs, we find $R_{\rm GCS}/R_{\rm e} = 0.6 \pm 0.1$. In Figure \ref{fig:mass_segregation}, GCs in low-mass ACSVCS (ACS Virgo Cluster Survey; \citealt{Cote2004}) galaxies with $M_z>-19$ show no overall significant mass segregation, and we find the same result when checking individual ACSFCS or ACSVCS galaxies. This suggests that the mass segregation observed in FCC~224 is unusual. We also note that if FCC~224 were observed at the distance of ~$\sim$~100 Mpc, only its bright GCs would be detected, and it would resemble one of the outliers in Figure~\ref{fig:rgc_re_mass} with a compact GC system.

It is difficult to compare the mass--radius trend of FCC~224 to DF2 and DF4, because there are not complete catalogs of those galaxies' GCs available, but they do appear different from FCC~224 in having many bright GCs located at large radii. The one other case where GC mass segregation was noticed in a UDG was NGC~5846-UDG1 \citep{Bar2022}. This UDG has an overall $R_{\rm GCS}/R_{\rm e} = 0.8 \pm 0.1$ \citep{Danieli2022}, while the median radius of the brightest 8 GCs is $0.5 R_{\rm e}$ (see Figure~2 of \citealt{Bar2022}). Both of these properties are very similar to FCC~224, although the NGC~5846-UDG1 has many more GCs in total ($54 \pm 9$).

\subsection{GC stellar population}
\label{sec:gc_stellar_pop}

\begin{figure}
\centering
\includegraphics[width=0.473\textwidth]{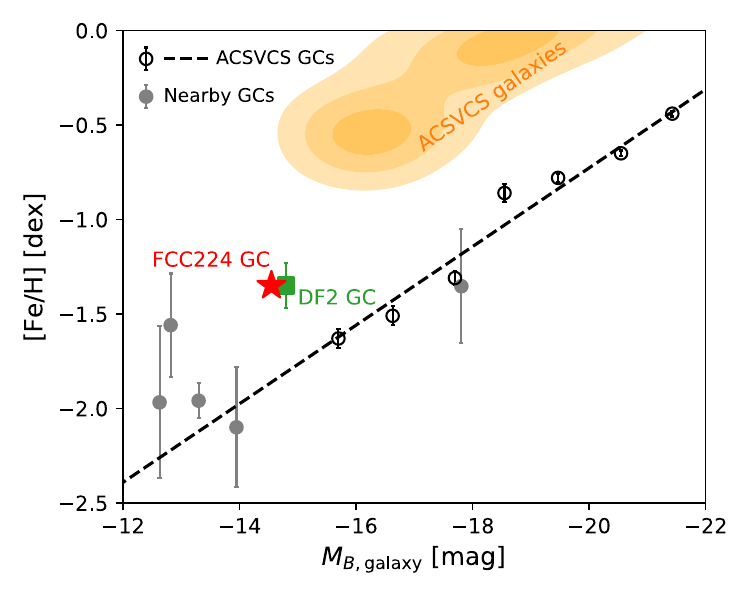}
\caption{Relation between mean metallicity [Fe/H] and host galaxy luminosity $M_B$. Virgo cluster galaxy GCs from ACSVCS are shown with black open dots and black dashed line), supplemented at the low luminosity end with nearby GCs (gray dots). FCC~224 (red star), DF2 (green square) GCs are overplotted. The orange contours show the distribution of the metallicity and luminosity of the ACSVCS galaxies themselves. The GCs in FCC~224 appear to be unusually metal-rich, with a metallicity similar to the host galaxy rather than being more metal-poor as is typical in other cases. In this respect, DF2 is very similar to FCC~224. FCC~224 itself is 0.1 dex higher than its GC system in this figure.}
\label{fig:feh_mb}
\end{figure}

\begin{figure*}
\centering
\includegraphics[width=0.9\textwidth]{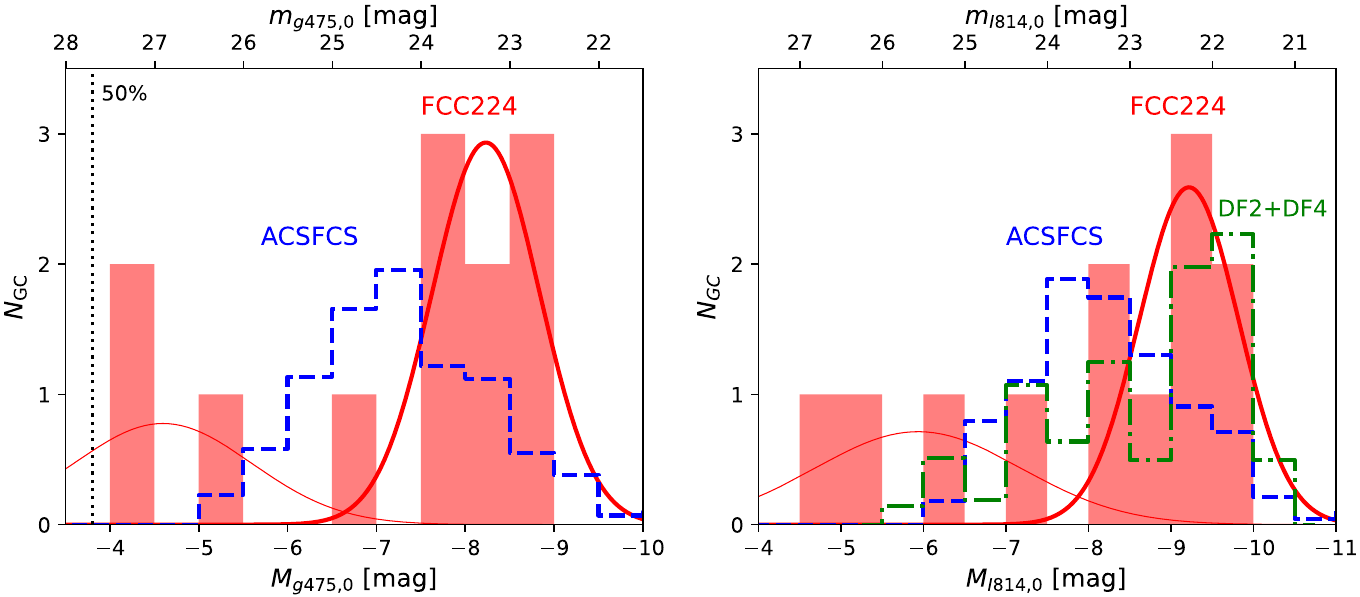}
\caption{The FCC~224 GCLFs in $g_{475}$ (left) and $I_{814}$ (right; red solid histograms). The two fitted Gaussian components for bright and faint peaks are represented by thick and thin red solid curves, respectively. The stacked GCLFs of ACSFCS low-mass galaxies ($M_z>-19$, blue dashed histograms) and DF2 and DF4 (green dot-dashed histograms) are overplotted, and they are normalized to have the same total GC number as FCC~224. The vertical dotted line is the 50\% completeness limit in $g_{475}$. FCC~224 has a top-heavy GCLF with few fainter GCs. The turnover magnitude is $\sim$~1 mag brighter than in other Fornax galaxies, and 0.2 mag fainter than in DF2 and DF4. The width of the main peak of the FCC~224 GCLF is also unusually narrow compared to other Fornax galaxies.}
\label{fig:gclf}
\end{figure*}

As seen in the CMD in Figure \ref{fig:gc_cmd}, the FCC~224 GCs have a strikingly narrow range of $g_{475}-I_{814}$ color. The brightest 7 GCs with $g_{475,0}<24$ have almost the same color of $0.73 \pm 0.02$ mag, only 0.04 mag bluer than the diffuse starlight of FCC~224. The color spread of 0.03 for the GCs overall is tiny compared to the median value of the low-mass ACSFCS and ACSVCS galaxies with $M_z>-19$ (0.24 and 0.27 mag, respectively in ACS F475W$-$F850LP), but is comparable to the ``monochromaticities'' of DF2 and DF4 \citep{vanDokkum2022b}. Some GC-rich UDGs show similarly narrow GC color distributions (DGSAT~I and NGC~5846-UDG1; \citealt{Janssens2022,Danieli2022}), while others do not (e.g., \citealt{Saifollahi2022}). 

Assuming the FCC~224 GCs have the same age as the diffuse galaxy light, we use their colors to estimate their corresponding metallicities [M/H]~$\simeq -1.35$ dex based on the {\tt FSPS} models. In Figure \ref{fig:feh_mb}, we show a relation between the mean metallicities of GC systems and their host galaxy luminosities based on the ACSVCS galaxies \citep{Peng2006}, with the low luminosity end supplemented by other nearby galaxies from \cite{Larsen2018}. We estimate the metallicities of the ACSVCS galaxies with the same color--metallicity relation in \cite{Peng2006}, using the photometry in \cite{Ferrarese2006}, and overplot the metallicity distribution of those galaxies (orange contours) in addition to their GCs. The metallicities of the ACSVCS galaxies are typically $\sim 0.7$ dex higher than their GCs. To put FCC~224 on the same plot, we estimate $M_B \approx -14.55$ and assume $[\alpha/{\rm Fe}]=0$ for the GCs. Its GCs appear to be unusually metal-rich, with a mean metallicity similar to the host galaxy (only $\sim 0.1$ dex difference), rather than being much more metal-poor as is typical for other galaxies. DF2 has a similar galaxy luminosity and GC metallicity to FCC~224 \citep{vanDokkum2018b}. DF4 also has a similar luminosity and almost identical GC colors (thus likely similar metallicity) to DF2 \citep{vanDokkum2022b}.

\subsection{GC luminosity function}
\label{sec:gclf}

\begin{figure}
\centering
\includegraphics[width=0.473\textwidth]{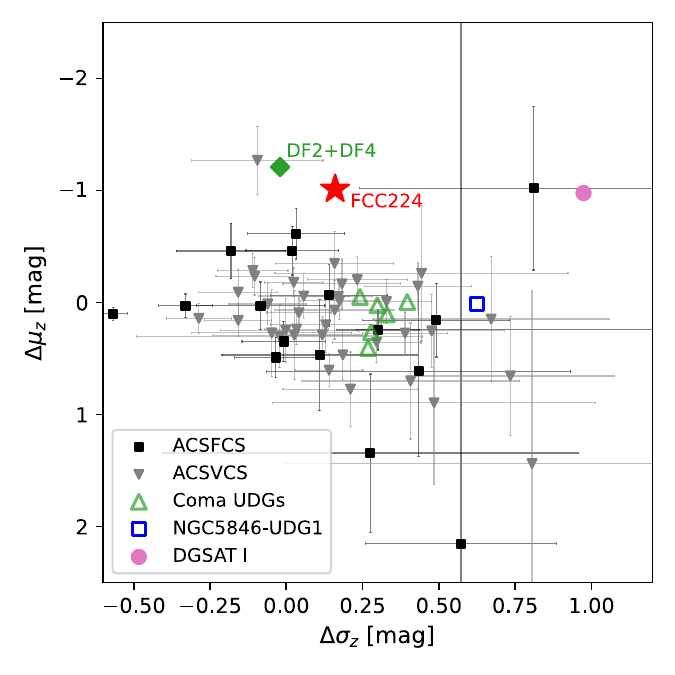}
\caption{The distribution of the GCLF turnover magnitude $\mu$ and width $\sigma$ in the ACS $z_{850}$ filter, relative to the trends with host galaxy luminosity (indicated by $\Delta$). Shown are FCC~224 (red star), DF2 and DF4 stacked (green diamond), other UDGs in the literature (green triangles, blue square and pink circle), low-luminosity ACSFCS (black squares) and ACSVCS galaxies (gray downward triangles) with $M_z>-19$ from \cite{Villegas2010}. The data points for FCC~224, DF2, and DF4 only represent the brighter component in the double-Gaussian fits from this work and \cite{Shen2021a}, while DGSAT~I, ACSFCS, and ACSVCS galaxies were fitted by single Gaussians. FCC~224 GCs are overluminous compared to almost all other galaxies, after considering the dependence of GCLF brightness and width on the host galaxy luminosity.}
\label{fig:gclf_mu_sigma}
\end{figure}

Figure \ref{fig:gclf} shows the GCLFs of FCC~224 in both of our {\it HST} filters. The background contamination has been subtracted. The GCLFs in both bands have a top-heavy shape, with $\sim 8$ GCs clustering at the bright end. Since there is also a fainter tail consisting of $\sim 4$ GCs, the GCLFs are better modeled with two Gaussians in order to measure the turnover magnitude, with one for the bright and narrow peak and the other for the faint and extended tail. Such double Gaussian fitting was also adopted for DF2 and DF4 for the same reason \citep{Shen2021a}. We plot both fitted Gaussian components in Figure \ref{fig:gclf}. The fainter peaks have mean and standard deviation values of $-4.6$ and 1.0 mag in $g_{475}$, and $-5.9$ and 1.3 mag in $I_{814}$. We note that the faint component is only designed to improve the fit, and we will not discuss it further. Based on the fit of the brighter peak, we find the GCLF turnover absolute magnitudes of about $-8.2$ in $g_{475}$ and $-9.2$ in $I_{814}$, while the best-fitted dispersions are about 0.61 and 0.60 mag for the two bands, respectively.

We overplot in Figure \ref{fig:gclf} the stacked GCLFs of the Fornax galaxies obtained with the ACSFCS GC candidates in \cite{Jordan2015}, including only the sources with a probability $p > 0.5$ of being GCs, and with host galaxies fainter than $M_{\rm F850LP}=-19$~mag. The ACS F475W filter was used in \cite{Jordan2015}, which is close enough to our $g_{475}$ band for a direct comparison. For $I_{814}$, we transform from the ACS F475W and F850LP magnitudes using {\tt FSPS}. Applying the Fornax distance of 20.0 Mpc \citep{Blakeslee2009}, we find that the Fornax GCLF has turnover magnitudes of $-7.25$ and $-8.18$ in the $g_{475}$ and $I_{814}$ bands, with dispersions of 0.94 and 0.93 mag, respectively (where the completeness limits are much deeper than the turnover magnitudes; \citealt{Villegas2010}). The GCLF of FCC~224 is $\sim 1$ mag brighter in turnover magnitude and narrower in dispersion compared to the Fornax GCLF. Note that we have measured the SBF distance of FCC~224 in Section \ref{sec:sbf}, so such a top-heavy GCLF should not result from assuming the wrong distance. If the turnover magnitude were normal, FCC~224 would have to be at a distance of $\sim 12.5$ Mpc, disagreeing at the $\sim 2 \sigma$ level with our SBF distance, and with the apparent association with the Fornax cluster based on position and redshift.

Moving beyond stacked GCLFs, we check if any individual galaxies in ACSFCS and ACSVCS are known to have similar top-heavy and narrow GCLFs to FCC~224. Figure \ref{fig:gclf_mu_sigma} illustrates the distribution of the mean magnitudes $\mu$ and dispersions $\sigma$ of the GCLFs in the ACS F850LP filter. The F850LP magnitudes for the plotted galaxies without observation in this filter are transformed from other filters using {\tt FSPS} models. These are based on single-Gaussian fits, with none of the GCLFs showing fainter tails similar to FCC~224, DF2, and DF4 that would require double-Gaussian functions. Since brighter galaxies tend to have brighter turnover magnitudes and larger dispersions, we show $\Delta \mu$ and $\Delta \sigma$, which mean the values after removing the dependences on host galaxy luminosity from \cite{Villegas2010}.  These galaxies show a trend for the GCLFs to be narrower as they are brighter. Even, so FCC~224 stands out with its bright GCLF peak. VCC~1886 is the one exception, with a GCLF that is apparently brighter and narrower than in FCC~224. However, we caution that in a large sample of galaxies, uncertainties in background corrections could cause a few cases to scatter to erroneously extreme values -- particularly for VCC~1886 which has only $\sim$~5 GCs. The result for FCC~224 is more secure because the bright GCs are found in a small region around the galaxy center, and because most of the GCs even have redshift confirmations. We also perform a Kolmogorov--Smirnov test between the FCC~224 GCLF and a Gaussian model with the mean and sigma predicted from \cite{Villegas2010} at the luminosity of FCC~224. The returned $p$-value is 0.03, indicating that the top-heavy GCLF of FCC~224 is highly unusual.

It is worth noting that \cite{Saifollahi2024} reported a secondary bright peak in addition to the normal distribution for their stacked sample of Fornax dwarfs from {\it Euclid}. The bright peak is close to the turn-over magnitude of FCC~224, and was traced to come mainly from three specific dwarfs in their sample. Further study of those galaxies individually, and of a larger sample of Fornax dwarfs, would be valuable.

The dwarf galaxies established as having similarly unusual GCLFs to FCC~224 are DF2 and DF4 \citep{vanDokkum2018b,vanDokkum2019,Shen2021a}. \cite{Shen2021a} reported that the combined GCLF of DF2 and DF4 has a brighter turnover magnitude $M_{\rm F606W,ACS} \approx -9$ and a dispersion of 0.4~mag, as well as a broader population of fainter GCs (as also seen in FCC~224). Given the homogeneous GC colors of ${\rm (F606W-F814W)_{ACS}} \approx 0.40$ measured by \cite{vanDokkum2022b} in DF2 and DF4, the implied ACS F814W turnover magnitude is about $-9.4$, which is almost the same as our FCC~224 result, as seen in the right panel of Figure~\ref{fig:gclf}. Compared to DF2 and DF4, the FCC~224 GCLF has a very similar shape, with its turnover magnitude only $\sim 0.2$ mag fainter and a slightly wider dispersion (see also Figure~\ref{fig:gclf_mu_sigma}).

We also show other GC-rich UDGs from the literature with their GCLFs measured \citep{Danieli2022,Janssens2022,Saifollahi2022} in Figure \ref{fig:gclf_mu_sigma}. Most of them are located in a very similar area to the ACSFCS and ACSVCS galaxies, which strengthens the impression that FCC~224, as well as DF2 and DF4, has an unusual GCLF. The exception is DGSAT~I (blue square at upper right), which has an overluminous GCLF and a broad peak from a single-Gaussian fit. Fitting the data with a two-component model would likely move this point to a similar region of the plot to FCC~224, DF2, and DF4. Also the Perseus UDGs R27 and W88 have clearly top-heavy GCLFs from the CMDs shown in \citet{Janssens2024}, which awaits further quantification. Interestingly, DGSAT~I and W88 have two of the most compact GC systems plotted in Figure~\ref{fig:rgc_re_mass}, which could be an indication of mass segregation, since faint outer GCs as in FCC~224 would not be detected in these distant systems.

\subsection{GC sizes}

\begin{figure*}
\centering
\includegraphics[width=0.95\textwidth]{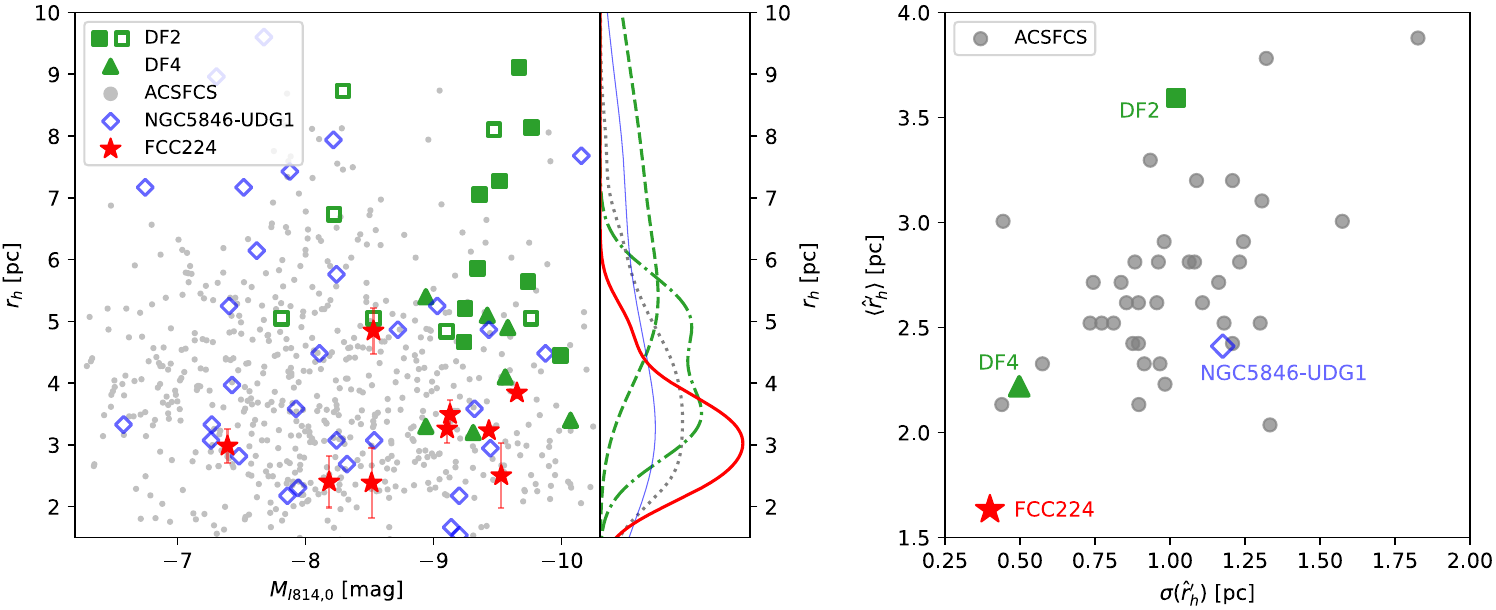}
\caption{{\it Left}: The distribution of GC size and $I_{814}$ absolute magnitude of FCC~224 (red stars with errorbars), DF2 (filled green squares from \citealt{vanDokkum2018b}, open green squares from \citealt{Trujillo2019}), DF4 (green triangles), low-mass ($M_z>-19$) ACSFCS galaxies (small gray dots), and NGC~5846-UDG1 (blue diamonds). The size distribution of each group of GCs is shown on the right side (red solid curve for FCC~224, green dashed curve for DF2, green dot-dashed curve for DF4, blue thin solid curve for NGC~5846-UDG1, and gray dotted curve for ACSFCS galaxies). {\it Right}: Mean GC size and size scatter for FCC~224, DF2, DF4, NGC~5846-UDG1, and each ACSFCS galaxy, corrected for the GC size dependence on the GC and host galaxy properties. The GCs in FCC~224 are over-compact, in contrast to DF2, DF4, NGC~5846-UDG1, and the other Fornax galaxies.}
\label{fig:gc_size}
\end{figure*}

As described in Section \ref{sec:gc_selection}, we obtain the half-light radius $r_h$ of the brightest nine GCs in FCC~224 using {\tt ishape}, shown in the left panel of Figure \ref{fig:gc_size}. The average $r_h$ of these GCs is 3.2 pc, with no noticeable correlation with brightness. All but one of them occupies a narrow size distribution near 3~pc. Although we are not able to measure the sizes of the three fainter GCs with this method, their $C_{3-7}$ concentrations imply that they are no larger than 5 pc. For comparison, most GCs of similar brightness in low-luminosity Fornax galaxies span a range of $r_h \sim$~3--5~pc (\citealt{Jordan2005} using the ACS/F850LP filter; see Figure~\ref{fig:gc_size}). Although the FCC~224 GCs fall within this range of sizes, they are biased low, with an average size of 3.2~pc compared to 3.8~pc for the other Fornax galaxies. To make sure that this difference is not caused by differing methodologies, we have measured the sizes of a random selection of 20 GCs from ACSFCS using our method, and find no systematic difference with \citet{Jordan2005}.

We next investigate whether or not any individual Fornax galaxies have GCs biased to small sizes as in FCC~224. \cite{Masters2010} reported that the mean GC size in each ACSFCS galaxy correlates with the luminosity, color, and local surface brightness of the galaxy, and also with the GC colors. This context needs to be considered for a fair comparison of FCC~224 GCs to those of other galaxies. To facilitate such comparisons, \cite{Masters2010} defined a new average GC size indicator $\hat{r}_h^{\prime}$ that removes the correlations with galaxy properties (see their Equations~7 and 8). Since the UDGs shown in Figure \ref{fig:gc_size} (FCC~224, DF2, DF4,NGC~5846-UDG1) are fainter than the ACSFCS galaxies, we have to extrapolate their equations to the lower luminosities to calculate $\hat{r}_h^{\prime}$. In the right panel of Figure \ref{fig:gc_size}, we show these corrected average sizes, as well as size scatter in each galaxy, $\sigma(\hat{r}_h^\prime$)\footnote{This quantity was not provided by \cite{Masters2010}, so we calculated it from the individual GC sizes.  We used the mean surface brightness of each galaxy provided in \cite{Masters2010} for their Eq.~(7), instead of the local surface brightness for each GC (which is not publicly available). Our $\hat{r}_h^\prime$ values calculated in this way agree well with those reported by \cite{Masters2010}.}. The $I_{814}$ magnitudes of ACSFCS galaxies are transformed from their ACS F475W and F850LP magnitudes based on {\tt FSPS} models. FCC~224 has a mean $\hat{r}_h^{\prime}$ of 1.6 pc, which is $\sim 35\%$ smaller than the typical value of 2.6 pc for ACSFCS galaxies. Another way to state this result is that the uncorrected mean GC size expected for FCC~224 is around 5~pc, in contrast to the 3.2 pc observed. It also has an unusually small corrected size scatter of 0.4~pc compared to $\sim$~0.8--1.2~pc for most of the other galaxies. Considering both mean size and scatter jointly, FCC~224 has more compact GCs overall than any other Fornax galaxy.

Figure~\ref{fig:gc_size} also shows GC sizes for DF2 and DF4, where again we have cross-checked the literature sizes using our own methods. The distances adopted here are 21.7 and 20.0 Mpc for DF2 and DF4, respectively \citep{Danieli2020,Shen2021b,Shen2023}. The mean sizes of the bright, spectroscopically confirmed GCs in DF2 and DF4 are larger than in FCC~224, with 6.7 pc and 4.1 pc, respectively (filled green square and triangles; \citealt{vanDokkum2018b,vanDokkum2019}). \cite{Trujillo2019} measured some of the fainter GC candidates in DF2 (shown in Figure \ref{fig:gc_size} as open green squares), and we see that these are similarly large to the brighter GCs. After corrections, the DF4 GC sizes are within the relation between mean size and scatter for Fornax galaxies, while DF2 is an outlier to larger sizes.

Besides these galaxies, we are aware of only one other UDG in the literature with its GC sizes analyzed in detail: NGC~5846-UDG1 \citep{Muller2021}. The bright, spectroscopically confirmed GCs in this galaxy are relatively compact on average, like FCC~224, but show a very large scatter, from $\sim$~1 to 7~pc, as Figure \ref{fig:gc_size} shows, adopting a distance of 26.5 Mpc \citep{Danieli2022}. The fainter GC candidates have more ``normal'' sizes. DGSAT~I is at a prohibitively large distance for secure measurements of GC sizes, but there are indications that they are unusually large \citep{Janssens2022}.

\section{Discussion} \label{sec:discussion}

In this section, we discuss the various properties of FCC~224 and its GC system presented in Section \ref{sec:results} and the possible formation and evolution mechanisms of this unusual dwarf galaxy. We have confirmed that the GC system is overluminous, with the GCLF turn-over magnitude $\sim 1$ magnitude brighter than in other Fornax galaxies, and missing most of the expected fainter GCs. We also found several other distinctive properties: the GC system is mass-segregated; the GCs have colors that are close to the host galaxy color, and with a narrow spread; the GCs have more compact sizes on average than in other dwarfs, and with a relatively narrow spread; and the galaxy's SFH is very close to a single burst, with no significant color gradient.

FCC~224 is remarkably similar to DF2 and DF4 in galaxy size and luminosity, and in the colors, magnitudes, and mass fraction ($\sim 3\%$ for both DF2 and DF4) of its GC system. However, all three galaxies differ noticeably in GC sizes and spatial distribution. A measurement of the DM content of FCC~224 is still needed to establish if this galaxy is a true analog of DF2/DF4, or if the similar GC properties are the product of convergent evolution. In the meantime, we can consider various formation scenarios for FCC~224, particularly those that have been proposed for DF2/DF4.

In the first general category of models, the GCs were born overluminous. The initial star cluster mass function is thought to be governed by an interplay between gas density and feedback \citep{Trujillo-Gomez2019,Andersson2024}. The case of DF2 was considered in detail by \cite{Trujillo-Gomez2021}, who used the present-day GCLF to constrain the initial physical conditions and cluster mass function. They found that the predominantly overluminous GCs required extremely dense and sheared conditions -- most likely in a gas-rich major merger. The starburst from this event could plausibly provide enough feedback to expand the galaxy into a UDG. This type of model would presumably work as well for FCC~224 as for DF2. The colors of stars and GCs in both galaxies provide new constraints.  While the GC monochromaticity might be the result of a single starburst followed by rapid quenching, the similar GC and host galaxy colors seem more problematic for a merger scenario. The galaxy would have a more extended SFH (including pre-merger stars) than the GCs, and some fine-tuning may be required to yield the same mean color, and no color gradient for the galaxy.

An extreme version of the major merger scenario is a ``bullet dwarf'' collision, where the impact speed is high enough for the progenitor galaxies to move apart after the collision and leave behind a new, dark matter-free galaxy that formed from shocked gas \citep{Silk2019}. This event would very naturally explain several of the observations of DF2, DF4, and FCC~224: the overluminous GCs (from the high-pressure and highly-sheared collision), the nearly uniform colors of the GCs and host galaxy, the short star formation timescale, and the diffuse nature of the galaxies (see also discussion in \citealt{Tang2024}). The latter two aspects would be consequences of the lack of the usual dark matter halo. Without the gravitational binding energy from a usual halo, stellar feedback can easily expel all the gas and quench the star formation, while the associated change in gravitational potential leads to an expansion of the stellar component.

A measurement of the dark matter content of FCC~224 will naturally be key for testing the bullet dwarf scenario. The surrounding regions of Fornax can also be canvassed to identify a possible ``twin'' to FCC~224 that was also born in the collision. For example, FCC~240 is located 13~arcmin away ($\sim$75 kpc in projection at the Fornax distance) and at first glance resembles FCC~224 in size, shape, and orientation \citep{Paudel2023}. We have measured its color from DECaLS, and it is only $\sim 0.05$~mag redder than FCC~224. However, it is $\sim 0.5$~mag brighter, shows twisted and irregular isophotes, and also appears to have a more normal GCLF. One would also expect that after a collision many Gyr ago, the pair of galaxies would have likely moved much farther apart.

The GC population of FCC~224 appears to exhibit radial mass segregation, with the GCs closer to the center being brighter than the outer ones. This may be because massive GCs are more sensitive to dynamical friction, which results in more orbital decay. To achieve strong mass segregation through dynamical friction, FCC~224 should not be too massive within $\sim 2R_{\rm e}$. DM-poor galaxies could reasonably achieve this, whether they are truly DM-free or have low-density DM halos (e.g., \citealt{Trujillo-Gomez2022,Liang2024}), although for DF2 and DF4, no significant mass segregation has been observed. Alternatively, the mass segregation could have been imprinted around the birth epoch of the clusters (e.g., \citealt{Elmegreen2024}). Theoretical modeling is needed of the mass segregation in FCC~224 and how it relates to the GCLF.

The unusually small GC sizes in FCC~224 seem difficult to explain in any DM-deficient model (whether through a bullet-dwarf collision or extreme stellar feedback). Although GCs formed in environments with high density and high star formation efficiency are likely to be more compact \citep{Choksi2021}, they should inflate and spontaneously reach a uniform size independent of their initial values through two-body relaxation and tidal shocks \citep{Gieles2010,Gieles2016}. To produce more compact GCs, a deeper external tidal field is needed, so that the GC outskirts are stripped and reach smaller sizes. This requirement of a deep gravitational potential appears at odds with DM deficiency.

The small GC sizes suggest another possible scenario where a strong tidal field preferentially disrupts the low-mass GCs and skews the GCLF to be top-heavy. Considering that the FCC~224 GCs are similar in mass, size, and galactocentric radius to the GCs in the centers of giant galaxies like the Milky Way, we can make an order of magnitude estimate that FCC~224 has a similar mass within $\sim 1$ kpc radius. However, given the very low stellar density of FCC~224, it would need to contain a huge amount of DM, $\gtrsim 100$ times more massive than the stars in the central region, allowing it to have a sufficiently deep gravitational potential. This mass ratio is plausible for a dwarf galaxy as a whole, out to its virial radius, but not for its central regions.

The second general category of models for the top-heavy GCLF is {\it evolution}. One possibility is that the initial mass function was fairly normal, but for some unexplained reason, all of the lower-mass clusters were disrupted.  However, as explained by \cite{Trujillo-Gomez2021}, this would lead to more field stars in the galaxy than observed.

An alternative scenario is that the low-mass GCs have merged into more massive systems at inner radii. \cite{DuttaChowdhury2020} estimates that the GC merger rate for DF2 is 0.03 Gyr${}^{-1}$. Given that the velocity dispersion of FCC~224 should be comparable to or greater than DF2, such GC mergers would not happen much more frequently. However, if FCC~224 has a cored dark matter halo, the GCs might congregate at a specific radius after undergoing a period of inflating by dynamical friction, known as core stalling (e.g., \citealt{DuttaChowdhury2019}). The GCs appear to merge more easily in this case as they are closer to each other, although numerical simulations of dwarfs with similar mass to FCC~224 suggests the merger rate is still too low to make a significant change to the GCLF \citep{Modak2023}. Even so, the radial positions of the luminous GCs in FCC~224 are suggestive of core stalling, and further theoretical work is needed, along with direct observational constraints on the DM content.

In summary, none of the scenarios discussed above seems to explain all the properties of FCC~224. Especially, there is a lack of theoretical studies in the literature on compact star clusters in dwarf galaxies. Since both the bullet dwarf and the tidal disruption models predict extreme DM properties, future exploration of FCC~224's DM halo is crucial. 

Also, the sparse observations of UDG GC systems so far are beginning to reveal a wide range of properties that deviate from the standard relations established for classical dwarfs. Considering FCC~224, DF2, DF4, NGC~5846-UDG1, and DGSAT~I, each of them shows a distinctly different combination of GCLF, GC sizes, GC radial distribution, and galaxy velocity dispersion -- and the causal connections between these properties are entirely unclear. A larger, systematic study of UDG GC systems in detail is needed to elucidate the patterns and to understand the underlying physics.

\section{Conclusion} \label{sec:conclusion}

In this work, we conduct a comprehensive analysis of an unusual LSB dwarf galaxy in the Fornax Cluster, FCC~224, and its GC system using new {\it HST} imaging and Keck/KCWI spectroscopy, along with existing DECaLS and {\it WISE} imaging. Our main findings and conclusion are as follows:

\begin{itemize}

\item 
The FCC~224 distance estimated through surface brightness fluctuations is $18.6 \pm 2.7$ Mpc, consistent with the Fornax Cluster distance of 20.0 Mpc.

\item 
FCC~224 is an old and metal-poor dwarf with a mean stellar age of $10.1_{-1.0}^{+1.3}$ Gyr and metallicity [M/H] of $-1.25_{-0.07}^{+0.05}$ dex. The galaxy does not show a significant color gradient, and the GCs have colors very close to the diffuse starlight with an extremely narrow color spread, suggesting an intense, single-burst star formation history.

\item
FCC~224 has an unusual GC system, with a very small color spread, radial mass segregation, a top-heavy GCLF, and systematically small sizes. 

\item
FCC~224 shows both similarities and differences in various properties compared to the DM-deficient galaxies NGC1052-DF2 and -DF4. We confirm that FCC~224 is a new case with a top-heavy GCLF similar to DF2 and DF4. Also, all these three dwarfs are UDGs and have monochromatic GCs. However, all three galaxies appear very different from each other in their GC sizes and spatial distributions.

\end{itemize}

We suggest that more theory work is needed in the future for galaxies like FCC~224. We have not found a known scenario that can fully fit the observations of FCC~224, particularly to explain the top-heavy GCLF and the unusually small GCs at the same time. The most crucial observation for FCC~224 in the near future for discriminating between models will be the dynamical mass measurement, as we find most of the models we have discussed have extreme dark matter properties. Moreover, in upcoming deep and wide-field surveys, more galaxies similar to FCC~224, DF2, and DF4 are expected to be found and analyzed, so that we will have a better understanding of this special population of galaxies.

\begin{acknowledgments}

We are grateful to the anonymous referee for numerous suggestions that clarified our results. Supported by National Science Foundation grant AST-2308390. Support for Program number HST-GO-17149 was provided through a grant from the STScI under NASA contract NAS5-26555. Data presented herein were obtained at the W.~M.\ Keck Observatory from telescope time allocated to the National Aeronautics and Space Administration through the agency’s scientific partnership with the California Institute of Technology and the University of California. The Observatory was made possible by the generous financial support of the W.~M.\ Keck Foundation. The authors wish to recognize and acknowledge the very significant cultural role and reverence that the summit of Maunakea has always had within the indigenous Hawaiian community. We are most fortunate to have the opportunity to conduct observations from this mountain. This research was supported by the Australian Research Council Centre of Excellence for All Sky Astrophysics in 3 Dimensions (ASTRO 3D), through project number CE170100013. AFM has received support from RYC2021-031099-I and PID2021-123313NA-I00 of MICIN/AEI/10.13039/501100011033/FEDER,UE, NextGenerationEU/PRT.

\end{acknowledgments}

\bibliography{fcc224}{}
\bibliographystyle{aasjournal}

\end{document}